\documentclass[preprint,prd,aps,showpacs,showkeys,nofootinbib]{revtex4}
\usepackage{graphicx}
\usepackage{dcolumn}
\usepackage{bm}
\usepackage{color}
\usepackage[dvipsnames]{xcolor}
\usepackage{amssymb}

\definecolor{light-gray}{gray}{0.78}
\definecolor{mid-gray}{gray}{0.55}
\definecolor{dark-gray}{gray}{0.32}

\begin{document}

\title{Higgs boson decays $h\rightarrow Z \gamma$ and $h\rightarrow m_V Z$ in the $U(1)_X$SSM}
\author{Xi Wang$^{1,2}$\footnote{wx$\_$0806@163.com}, Shu-Min Zhao$^{1,2}$\footnote{zhaosm@hbu.edu.cn}, Tong-Tong Wang$^{1,2}$, Lu-Hao Su$^{1,2}$, Wei Li$^{1,2}$, Ze-Ning Zhang$^{1,2}$, Zhong-Jun Yang$^{3}$, Tai-Fu Feng$^{1,2,3}$}

\affiliation{$^1$ Department of Physics, Hebei University, Baoding 071002, China}
\affiliation{$^2$ Key Laboratory of High-precision Computation and Application of Quantum Field Theory of Hebei Province, Baoding 071002, China}
\affiliation{$^3$ Department of Physics, Chongqing University, Chongqing 401331, China}
\date{\today}

\begin{abstract}
We present a detailed analysis of the Higgs boson decays $h\rightarrow Z \gamma$ and $h\rightarrow m_V Z$ in the $U(1)_X$SSM, with $m_V$ denoting one of the mesons ($\omega,\rho,\phi,J/{\psi},\Upsilon$). Using the effective Lagrangian method, we calculate the effective constants $C_{\gamma Z}$ (CP-even) and $\tilde{C}_{\gamma Z}$ (CP-odd) for the vertex $h \gamma Z$, which are corrected by the new particle loop diagrams. Numerically, the ratio ${\Gamma}_{U(1)_X}(h\rightarrow Z \gamma )/{\Gamma}_{SM}(h\rightarrow Z \gamma)$ is between $1.02-1.35$. For the vector mesons $\omega,\rho,\phi$ and $J/{\psi}$, the ratios ${\Gamma}_{U(1)_X}(h\rightarrow m_V Z )/{\Gamma}_{SM}(h\rightarrow m_V Z)$ are mainly distributed in the range of ($1.01-1.45$). When $m_V$  represents $\Upsilon$, the ratio ${\Gamma}_{U(1)_X}(h\rightarrow \Upsilon Z )/{\Gamma}_{SM}(h\rightarrow \Upsilon Z)$ is mostly located in the range of ($1.002-1.30$). The aim of this work is to provide a reference for probing the $U(1)_X$SSM model via the Higgs decays $h\rightarrow Z \gamma$ and $h\rightarrow m_V Z$.
\end{abstract}


\maketitle

\section{Introduction}

Since ATLAS and CMS discovered the Higgs boson $h$ in 2012 \cite{IN1,IN2}, people have put substantial effort to study the Higgs boson. The latest experimental data show that the measured mass of the Higgs boson is $m_h = 125.25 \pm 0.17 ~{\rm GeV}$ \cite{pdg2022}.

As a new elementary particle, $h$ is consistent with the neutral Higgs boson predicted by the Standard Model (SM) to a large extent. The interaction between Higgs boson and electroweak gauge bosons ($W$, $Z$, $\gamma$) is now well established and Higgs boson decays to $W W^*$, $Z Z^*$ and $\gamma \gamma$ have all been accurately measured \cite{WW1,ZZ5,rr7}. Although people do many works on decay process $h\rightarrow Z \gamma$ \cite{LHC1,LHC2,LHC3,LHC4}, there is still no evidence for the existence of this decay process. With the Higgs boson mass $m_h = 125.09~{\rm GeV}$, the ATLAS Collaboration has found that the upper limit on the production cross section times the branching ratio for $pp \rightarrow h \rightarrow Z \gamma$ is 6.6 (5.2) times the SM prediction at the 95\% confidence level \cite{FJ1,FJ2,FJ3,FJ4}. There is no $h \gamma Z$ coupling at tree level, but it can be produced by loop diagrams \cite{IN3,IN4}. This coupling is very important for exploring the new physics.

These authors investigated the process $h\rightarrow m_V \gamma$ in considerable detail, with $m_V$ denoting a meson \cite{FA1,FA2,FA3}. Because the on-shell photon is massless and has no longitudinal polarization, $m_V$ is only a transversely polarized vector meson for the decay process $h\rightarrow m_V \gamma$. With the phenomenological Lagrangian, the exclusive weak radiative Higgs decays $h\rightarrow m_V V(V=Z,W)$ are examined as probes for nonstandard couplings \cite{IN8}.
Based on decay topologies, their contributions are classified into two groups: direct contributions and indirect contributions. For the direct contributions, the quarks making up the meson directly couple to the Higgs boson. For the indirect contributions, the meson is converted by an off-shell vector boson through the local matrix element \cite{IN5,IN6}. The decay process $h\rightarrow m_V \gamma$ is strongly disturbed by both direct and indirect contributions \cite{FA1,FA2,FA3}. The indirect contributions of the decay $h\rightarrow m_V Z$ are generated from the effective $h \gamma Z$ vertex, which are more important than the direct contributions, especially when $m_V$ is a light vector meson \cite{IN8}. It is proposed in the Refs.\cite{QCD1,QCD2,QCD3,QCD4} that the QCD factorization is used for the rare weak radiative Higgs boson decays $h\rightarrow m_V Z$.

The $U(1)_X$SSM is the $U(1)$ expansion of MSSM, whose local gauge group is $SU(3)_C\times SU(2)_L \times U(1)_Y\times U(1)_X$ \cite{Sarah1,Sarah2,Sarah3}. On the basis of the MSSM, three Higgs singlets and right-handed neutrinos are added. Therefore, light neutrinos obtain tiny masses through the seesaw mechanism, which can explain the results of neutrino oscillation experiment. In the $U(1)_X$SSM \cite{sm2}, the little hierarchy problem in MSSM is relieved by the right-handed neutrinos, sneutrinos and additional Higgs singlets. The $S$ field after vacuum spontaneous breaking can alleviate the $\mu$ problem of MSSM. In this work, we study the Higgs boson decays $h\rightarrow Z \gamma$ and $h\rightarrow m_V Z$ with $m_V$ denoting $\omega,\rho,\phi,J/{\psi}$ and $\Upsilon$ in the $U(1)_X$SSM, and we briefly discuss the numerical values of the processes $h\rightarrow \gamma \gamma$ and $h\rightarrow V V^*(V=Z,W)$. The relevant Feynman diagrams are derived and numerically analyzed. From the numerical results, we obtain reasonable parameter space.

In the following, we introduce the specific form of $U(1)_X$SSM and its superfields in Sec.II. We give related formulas of the Higgs boson decays $h\rightarrow Z \gamma$ and $h\rightarrow m_V Z$ in Sec.III. The input parameters and numerical results are shown in Sec.IV. The last section is used for the discussion and conclusion. Finally, some couplings are collected in Appendix.A.

\section{The relevant content of $U(1)_X$SSM}
$SU(3)_C\times SU(2)_L \times U(1)_Y\times U(1)_X$ is the local gauge group of $U(1)_X$SSM, which is the $U(1)$ expansion of MSSM \cite{UU1,UU3,UU4}.
There are new superfields in $U(1)_X$SSM, such as three Higgs
singlets $\hat{\eta},~\hat{\bar{\eta}},~\hat{S}$ and right-handed neutrinos $\hat{\nu}_i$,
which are beyond the MSSM. Light neutrinos gain tiny masses at the tree level through the seesaw mechanism. The $U(1)_X$SSM is anomaly free \cite{UU3}.
It is necessary to consider loop corrections in order to get the 125 GeV Higgs boson mass \cite{LCTHiggs1,LCTHiggs2}. We can find the particle content and charge assignments for $U(1)_X$SSM in our previous work \cite{tt1}.

According to $U(1)_X$SSM, the superpotential is
\begin{eqnarray}
&&W=l_W\hat{S}+\mu\hat{H}_u\hat{H}_d+M_S\hat{S}\hat{S}-Y_d\hat{d}\hat{q}\hat{H}_d-Y_e\hat{e}\hat{l}\hat{H}_d+\lambda_H\hat{S}\hat{H}_u\hat{H}_d
\nonumber\\&&\hspace{0.6cm}+\lambda_C\hat{S}\hat{\eta}\hat{\bar{\eta}}+\frac{\kappa}{3}\hat{S}\hat{S}\hat{S}+Y_u\hat{u}\hat{q}\hat{H}_u+Y_X\hat{\nu}\hat{\bar{\eta}}\hat{\nu}
+Y_\nu\hat{\nu}\hat{l}\hat{H}_u.
\end{eqnarray}

There are two Higgs doublets and three Higgs singlets
\begin{eqnarray}
&&\hspace{1cm}H_{u}=\left(\begin{array}{c}H_{u}^+\\{1\over\sqrt{2}}\Big(v_{u}+H_{u}^0+iP_{u}^0\Big)\end{array}\right),
~~~~~~
H_{d}=\left(\begin{array}{c}{1\over\sqrt{2}}\Big(v_{d}+H_{d}^0+iP_{d}^0\Big)\\H_{d}^-\end{array}\right),
\nonumber\\&&\eta={1\over\sqrt{2}}\Big(v_{\eta}+\phi_{\eta}^0+iP_{\eta}^0\Big),~~~
\bar{\eta}={1\over\sqrt{2}}\Big(v_{\bar{\eta}}+\phi_{\bar{\eta}}^0+iP_{\bar{\eta}}^0\Big),~~
S={1\over\sqrt{2}}\Big(v_{S}+\phi_{S}^0+iP_{S}^0\Big).
\end{eqnarray}
The Higgs superfields $H_u$, $H_d$, $\eta$, $\bar{\eta}$ and $S$ each have their vacuum expectation values(VEVs), listed as $v_u,~v_d,~v_\eta$, $v_{\bar\eta}$ and $v_S$, respectively.
We define two angles as $\tan\beta=v_u/v_d$ and $\tan\beta_\eta=v_{\bar{\eta}}/v_{\eta}$. The $5\times5 $ mass squared matrix is formed when the neutral CP-even parts of
$H_u,~H_d,~\eta,~\bar{\eta}$ and $S$ mix together. Its lightest eigenvalue corresponds to the lightest CP-even Higgs mass.

The soft SUSY breaking terms  are shown as
\begin{eqnarray}
&&\mathcal{L}_{soft}=\mathcal{L}_{soft}^{MSSM}-B_SS^2-L_SS-\frac{T_\kappa}{3}S^3-T_{\lambda_C}S\eta\bar{\eta}
+\epsilon_{ij}T_{\lambda_H}SH_d^iH_u^j\nonumber\\&&\hspace{1cm}
-T_X^{IJ}\bar{\eta}\tilde{\nu}_R^{*I}\tilde{\nu}_R^{*J}
+\epsilon_{ij}T^{IJ}_{\nu}H_u^i\tilde{\nu}_R^{I*}\tilde{l}_j^J
-m_{\eta}^2|\eta|^2-m_{\bar{\eta}}^2|\bar{\eta}|^2-m_S^2S^2\nonumber\\&&\hspace{1cm}
-(m_{\tilde{\nu}_R}^2)^{IJ}\tilde{\nu}_R^{I*}\tilde{\nu}_R^{J}
-\frac{1}{2}\Big(M_S\lambda^2_{\tilde{X}}+2M_{BB^\prime}\lambda_{\tilde{B}}\lambda_{\tilde{X}}\Big)+h.c~.
\end{eqnarray}

We adopt $Y^Y$ to represent the $U(1)_Y$ charge and $Y^X$ to represent the $U(1)_X$ charge. $U(1)_Y$ and $U(1)_X$ are two Abelian groups that produce a new effect: the gauge kinetic mixing. This effect can also be induced through RGEs even if it is zero value at $M_{GUT}$.

Based on the fact that the two Abelian gauge groups are unbroken, we can do a basis conversion by using the rotation matrix $R$ ($R^T R=1$) \cite{UMSSM5,B-L1,B-L2,gaugemass} :
\begin{eqnarray}
&&D_\mu=\partial_\mu-i\left(\begin{array}{cc}Y^Y,&Y^X\end{array}\right)
\left(\begin{array}{cc}g_{Y},&g{'}_{{YX}}\\g{'}_{{XY}},&g{'}_{{X}}\end{array}\right)R^TR
\left(\begin{array}{c}A_{\mu}^{\prime Y} \\ A_{\mu}^{\prime X}\end{array}\right)\;,
\end{eqnarray}
here, the gauge fields of $U(1)_Y$ and $U(1)_X$ are expressed as $A_{\mu}^{\prime Y}$ and $A^{\prime X}_\mu$ respectively. We redefine the following
\begin{eqnarray}
&&\left(\begin{array}{cc}g_{Y},&g{'}_{{YX}}\\g{'}_{{XY}},&g{'}_{{X}}\end{array}\right)
R^T=\left(\begin{array}{cc}g_{1},&g_{{YX}}\\0,&g_{{X}}\end{array}\right)~,~~~~
R\left(\begin{array}{c}A_{\mu}^{\prime Y} \\ A_{\mu}^{\prime X}\end{array}\right)
=\left(\begin{array}{c}A_{\mu}^{Y} \\ A_{\mu}^{X}\end{array}\right)\;.
\end{eqnarray}
$g_X$ is the gauge coupling constant of the $U(1)_X$ group.
$g_{YX}$ is the mixing gauge coupling constant of  $U(1)_Y$ group and $U(1)_X$ group.
Then, the covariant derivatives of $U(1)_X$SSM can be written as
\begin{eqnarray}
&&D_\mu=\partial_\mu-i\left(\begin{array}{cc}Y^Y,&Y^X\end{array}\right)
\left(\begin{array}{cc}g_{1},&g_{{YX}}\\0,&g_{{X}}\end{array}\right)
\left(\begin{array}{c}A_{\mu}^{Y} \\ A_{\mu}^{X}\end{array}\right)\;.
\end{eqnarray}

The gauge fields $A^{X}_\mu,~A^Y_\mu$ and $V^3_\mu$ mix together at the tree level, and produce a $3\times3 $ mass squared matrix for neutral gauge bosons \cite{UU3}. We apply two mixing angles $\theta_{W}$ and $\theta_{W}'$ to diagonalize this matrix. $\theta_{W}$ is the Weinberg angle. As a new mixing angle, $\theta_{W}'$ is defined as
\begin{eqnarray}
\sin^2\theta_{W}'\!=\!\frac{1}{2}\!-\!\frac{[(g_{{YX}}+g_{X})^2-g_{1}^2-g_{2}^2]v^2+
4g_{X}^2\xi^2}{2\sqrt{[(g_{{YX}}+g_{X})^2+g_{1}^2+g_{2}^2]^2v^4+8g_{X}^2[(g_{{YX}}+g_{X})^2-g_{1}^2-g_{2}^2]v^2\xi^2+16g_{X}^4\xi^4}},
\end{eqnarray}
with $v=\sqrt{v_u^2+v_d^2}$ and $\xi=\sqrt{v_\eta^2+v_{\bar{\eta}}^2}$. We deduce the eigenvalues of the mass squared matrix for neutral gauge bosons. One is zero mass corresponding to the photon. The other two values are for $Z$ and $Z^{\prime}$ \cite{UU4,tt1}.

In the basis $(H_d^-,H_u^{+,*})$ and $(H_d^{-,*},H_u^+)$, the definition of the  mass squared matrix for charged Higgs is given by
\begin{eqnarray}
m^2_{H^-}= \left(
\begin{array}{cc}
m_{H_d^- H_d^{-,*}} &m^*_{H_u^{+,*} H_d^{-,*}}\\
m_{H_d^- H_u^+} &m_{H_u^{+,*} H_u^+}\end{array}
\right),
 \end{eqnarray}
\begin{eqnarray}
&&m_{H_d^- H_d^{-,*}}=m^2_{H_d}+ \frac{1}{8} \Big[(g_{2}^{2} + g_{X}^{2}) v_{d}^{2}
+ (g_{2}^{2} - g_{X}^{2}) v_{u}^{2}+(g_{1}^{2} + g_{Y X}^{2}) ( v_{d}^{2}- v_{u}^{2})-2 g_{X}^{2} v_{\bar{\eta}}^{2}
\nonumber\\&&\hspace{1.8cm}+ 2 \Big((g_{Y X} g_{X}) (-v_{\bar{\eta}}^{2}-v_{u}^{2}+v_{d}^{2}+v_{\eta}^{2})+g_{X}^{2} v_{\eta}^{2}\Big)\Big]
\nonumber\\&&\hspace{1.8cm}+ \frac{1}{2} \Big(2 |\mu|^2+2 \sqrt{2} v_S \Re{(\mu {\lambda}^*_{H})}+v_S^2 |{\lambda}_{H}|^2\Big),
\\&&m_{H_d^- H_u^+}= \frac{1}{2} \Big[2 ({\lambda}_{H} l_{W^*}+B_{\mu})+{\lambda}_{H} (2 \sqrt{2} v_S M^*_S-v_{d} v_{u} {\lambda}^*_{H}+v_{\eta} v_{\bar{\eta}} {\lambda}^*_{C}+v_S^2 {\kappa}^*)\nonumber\\&&\hspace{1.8cm}+\sqrt{2} v_S T_{{\lambda}_{H}}\Big]+ \frac{1}{4} g_{2}^{2} v_{d} v_{u},
\\&&m_{H_u^{+,*} H_u^+}=m^2_{H_u}+ \frac{1}{8} \Big[(g_{2}^{2} - g_{X}^{2}) v_{d}^{2}
+ (g_{2}^{2} + g_{X}^{2}) v_{u}^{2}+(g_{1}^{2} + g_{Y X}^{2}) (v_{u}^{2} -v_{d}^{2})-2 g_{X}^{2} v_{\eta}^{2}
\nonumber\\&&\hspace{1.8cm}+ 2 \Big((g_{Y X} g_{X}) (v_{\bar{\eta}}^{2}+v_{u}^{2}-v_{d}^{2}-v_{\eta}^{2})+g_{X}^{2} v_{\bar{\eta}}^{2}\Big)\Big]
\nonumber\\&&\hspace{1.8cm}+ \frac{1}{2} \Big[2 |\mu|^2+2 \sqrt{2} v_S \Re{(\mu {\lambda}^*_{H})}+v_S^2 |{\lambda}_{H}|^2\Big].
\end{eqnarray}
This matrix is diagonalized by $Z^+$:
\begin{eqnarray}
Z^+ m^2_{H^-} Z^{+,\dagger} = m^{dia}_{2,H^-},
 \end{eqnarray}
with
\begin{eqnarray}
H_d^-=\sum\limits_{j} Z^+_{j1} H_j^-,~~~H_u^+=\sum\limits_{j} Z^+_{j2} H_j^+.
 \end{eqnarray}

Moreover, the chargino mass matrix, down type squark mass squared matrix, up type squark mass squared matrix, slepton mass squared matrix and CP-even Higgs mass squared matrix are needed during calculation. These mass matrices can be found in Refs.\cite{UU1,UU3}.

Here are some commonly used couplings. The CP-even Higgs bosons interact with charginos, whose explicit form reads as
\begin{eqnarray}
&&\mathcal{L}_{H {\chi}^{\pm} {\chi}^{\pm}}= -\frac{1}{\sqrt{2}} \Big(g_{2} U_{j1}^{*} V_{i2}^{*} Z_{k2}^H+U_{j2}^{*} (g_{2} V_{i1}^{*} Z_{k1}^H+{\lambda}_{H} V_{i2}^{*} Z_{k5}^H)\Big)P_L
\nonumber\\&&\hspace{1.9cm}-\frac{1}{\sqrt{2}} \Big(g_{2} U_{i1} V_{j2} Z_{k2}^H+U_{i2} (g_{2} V_{i1} Z_{k1}^H+{\lambda}^{*}_{H} V_{j2} Z_{k5}^H)\Big)P_R.\label{coupling1}
\end{eqnarray}
The specific form of $Z-{\chi}^{\pm}-{\chi}^{\pm}$ is as follows:
\begin{eqnarray}
&&\mathcal{L}_{Z {\chi}^{\pm} {\chi}^{\pm}}= \frac{1}{2} \Big(2 g_{2} U_{j1}^{*} \cos\theta_{W} \cos\theta_{W}' U_{i1}+U_{j2}^{*} (-g_{1} \cos\theta_{W}' \sin\theta_{W} \nonumber\\&&\hspace{1.4cm}+g_{2} \cos\theta_{W} \cos\theta_{W}'+(g_{Y X}+g_{X}) \sin\theta_{W}') U_{i2}\Big) {\gamma}_{\mu}P_L\nonumber\\&&\hspace{1.4cm}+ \frac{1}{2} \Big(2 g_{2} V_{i1}^{*} \cos\theta_{W} \cos\theta_{W}' V_{j1}+V_{i2}^{*} (-g_{1} \cos\theta_{W}' \sin\theta_{W} \nonumber\\&&\hspace{1.4cm}+g_{2} \cos\theta_{W} \cos\theta_{W}'+(g_{Y X}+g_{X}) \sin\theta_{W}') V_{j2}\Big) {\gamma}_{\mu}P_R.\label{coupling2}
\end{eqnarray}
In the above two equations, $P_L=\frac{1-{\gamma}_5}{2}$ and $P_R=\frac{1+{\gamma}_5}{2}$. The mass matrix of chargino is diagonalized by the rotation matrixes $U$ and $V$.

We also deduce the vertex of $Z-\tilde{L}_i-\tilde{L}^{*}_j$,
\begin{eqnarray}
&&\mathcal{L}_{Z\tilde{L}\tilde{L}^{*}}=\frac{1}{2}\tilde{e}^{*}_j
\Big[(g_2\cos\theta_W\cos\theta_W^\prime-g_1\cos\theta_W^\prime\sin\theta_W
+g_{YX}\sin\theta_W^\prime)\sum_{a=1}^3Z_{i,a}^{E,*}Z_{j,a}^E\nonumber\\&&\hspace{1.2cm}
+\Big((2g_{YX}+g_X)\sin\theta_W^\prime-2g_1\cos\theta_W^\prime\sin\theta_W\Big)
\sum_{a=1}^3Z_{i,3+a}^{E,*}Z_{j,3+a}^E\Big](p^\mu_{i}-p^\mu_j)\tilde{e}_iZ_{\mu}.
\end{eqnarray}
To save space in the text, the remaining vertexes can be found in the Appendix.A and Ref.\cite{UU3}.

\section{Analytical formula}

\begin{figure}[ht]
\setlength{\unitlength}{5.0mm}
\centering
\includegraphics[width=6.5in]{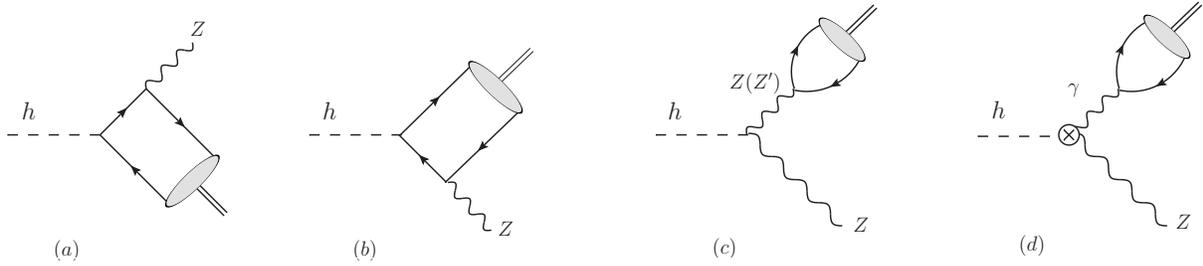}
\caption{ The major drawings that contribute to the decay $h\rightarrow m_V Z$.}\label{N1}
\end{figure}

The relevant formulas of decay processes $h\rightarrow Z \gamma$ and $h\rightarrow m_V Z$  are discussed in this section. The Fig.\ref {N1} depicts the appropriate Feynman diagrams for the Higgs boson weak hadronic decay $h\rightarrow m_V Z$. Fig.\ref {N1}(a) and Fig.\ref {N1}(b) show the direct contributions. Fig.\ref {N1}(c) and Fig.\ref {N1}(d) show the indirect contributions. The effective vertex $h\rightarrow Z {\gamma}^*$ from the one-loop diagrams is represented by the crossed circle in the last graph. According to the Ref.\cite{IN8}, for the direct contributions, the quarks making up the meson directly couple to the Higgs boson. For the indirect contributions, there is such a decay process $h\rightarrow Z Z^*/Z {\gamma}^*\rightarrow m_V Z$, where $Z^*/{\gamma}^*$ is an off-shell boson to form the final meson.
In fact, $Z'$ exchange appears in Fig.\ref {N1}(c) with the replacement of virtual $Z$. Considering the constraint of $Z'$ mass larger than 5.1 TeV which is very heavy, we neglect $Z'$
contribution. In the numerical analysis of the propagators,
$|\frac{1}{m^2_V-m^2_Z}| \sim \frac{1}{90^2}$ and  $|\frac{1}{m^2_V-m^2_{Z'}}| \sim \frac{1}{5100^2}$.
The latter is about $10^{-4}$ as much as the former and it can be ignored. So the exchange of $Z'$ is not considered in the calculation.

Among them, $h\rightarrow Z Z^*$ can occur at tree level in the SM. Although the vertex $h \gamma Z$ does not exist at tree level, it can be created by loop diagrams. The nonstandard $h\gamma Z$ vertex should be considered in the $U(1)_X$SSM. Here is the concrete expression of the effective Lagrangian for $h\gamma Z$:
\begin{eqnarray}
\mathcal{L}_{eff}= \frac{\alpha}{4 \pi v} \Big(\frac{2 C_{\gamma Z}}{\sin \theta_{W} \cos \theta_{W}} h F_{\mu v} Z^{\mu v}-\frac{2 {\tilde{C}}_{\gamma Z}}{\sin \theta_{W} \cos \theta_{W}} h F_{\mu v} {\tilde{Z}}^{\mu v}\Big).\label{f1}
\end{eqnarray}

The decay width of $h\rightarrow Z \gamma$ is deduced by using the effective Lagrangian in Eq.(\ref{f1}),
\begin{eqnarray}
\Gamma(h \rightarrow Z \gamma)= \frac{{\alpha}^2 m^3_h}{32 {\pi}^3 v^2 \sin^2\theta_{W} \cos^2 \theta_{W}}\Big(1-\frac{m^2_Z}{m^2_h}\Big)^3 (|C_{\gamma Z}|^2+|{\tilde{C}}_{\gamma Z}|^2).
\end{eqnarray}

The detailed derivation of $h\rightarrow m_V Z$ is shown in the Refs.\cite{FA1,FA2,FA3,IN8,sun4}. The decay width of $h\rightarrow m_V Z$ is shown as
\begin{eqnarray}
&&\hspace{-0.5cm}\Gamma(h\rightarrow m_V Z)= \frac{m^3_h}{4 \pi v^4} {\lambda}^{1/2}(1,r_Z,r_V)(1-r_Z-r_V)^2 \nonumber\\&&\hspace{2cm}{\times} \Big[|F^{VZ}_{\parallel}|^2+{\frac{8 r_Z r_V}{(1-r_Z-r_V)^2}} (|F^{VZ}_{\perp}|^2+|\tilde{F}^{VZ}_{\perp}|^2)\Big],\label{f4}
\end{eqnarray}
with $\lambda(x,y,z)=(x-y-z)^2-4 y z$, $r_Z=m^2_Z/m^2_{h}$ and $r_V=m^2_V/m^2_{h}$. For all mesons, the mass ratio $r_V=m^2_V/m^2_{h}$ is tiny, but it can make considerable contributions to the transverse polarization state for $h\rightarrow m_V Z$. Thus, we retain $r_V=m^2_V/m^2_{h}$ in order to obtain better results.

In Eq.(\ref{f4}), $F^{VZ}_{\parallel}$, $F^{VZ}_{\perp}$ and $\tilde{F}^{VZ}_{\perp}$ are divided into direct and indirect parts. The specific forms of indirect contributions are as follows:
\begin{eqnarray}
&&F^{VZ}_{\parallel indirect}= \frac{\kappa_Z}{1-r_V/r_Z} \sum_{q} f^q_V v_q+C_{\gamma Z} \frac{\alpha (m_V)}{4 \pi} \frac{4 r_Z}{1-r_Z-r_V} \sum_{q} f^q_V Q_q,
\nonumber\\&&F^{VZ}_{\perp indirect}= \frac{\kappa_Z}{1-r_V/r_Z} \sum_{q} f^q_V v_q+C_{\gamma Z} \frac{\alpha (m_V)}{4 \pi} \frac{1-r_Z-r_V}{r_V} \sum_{q} f^q_V Q_q,
\nonumber\\&&\tilde{F}^{VZ}_{\perp indirect}= \tilde{C}_{\gamma Z} \frac{\alpha (m_V)}{4 \pi} \frac{{\lambda}^{1/2}(1,r_Z,r_V)}{r_V} \sum_{q} f^q_V Q_q.\label{f5}
\end{eqnarray}
The vector and axial-vector couplings of $Z \bar{q} q$ are defined as $v_q =\frac{T^q_3}{2}-Q_q \sin^2\theta_{W}$ and $a_q =\frac{T^q_3}{2}$ respectively. The vector meson decay constant $f^q_V$ can be written as
\begin{eqnarray}
\Big<V(k,\varepsilon)|\bar{q} {\gamma}^{\mu} q| 0\Big>= -i f^q_V m_V {\varepsilon}^{* \mu},~~~q = u, d, s,...
\end{eqnarray}
We use the following relationship to get the results
\begin{eqnarray}
Q_V f_V=\sum_{q} Q_q f^q_V,~~~~\sum_{q} f^q_V v_q=f_V v_V.
\end{eqnarray}

The concrete forms of $C_{\gamma Z}$ and $\tilde{C}_{\gamma Z}$ in Eq.(\ref{f5}) read as
\begin{eqnarray}
&&C_{\gamma Z}=C^{SM}_{\gamma Z}+C^{U(1)_X}_{\gamma Z},~~~~\tilde{C}_{\gamma Z}=\tilde{C}^{SM}_{\gamma Z}+\tilde{C}^{U(1)_X}_{\gamma Z},
\nonumber\\&&C^{SM}_{\gamma Z}= \sum_{q} \frac{2 N_c Q_q v_q}{3} A_f({\tau}_q,r_Z)+\sum_{l} \frac{2 Q_l v_l}{3} A_f({\tau}_l,r_Z)- \frac{1}{2} A^{\gamma Z}_W({\tau}_W,r_Z),
\nonumber\\&&\tilde{C}^{SM}_{\gamma Z}= \sum_{q} \tilde{\kappa}_q N_c Q_q v_q B_f({\tau}_q,r_Z)+\sum_{l} \tilde{\kappa}_l Q_l v_l B_f({\tau}_l,r_Z),
\end{eqnarray}
with ${\tau}_i=4 m^2_i/m^2_{h}$. $C^{SM}_{\gamma Z}$ and $\tilde{C}^{SM}_{\gamma Z}$ are the contributions of SM  to $h\rightarrow Z \gamma$. $A_f$,~$B_f$ and $A^{\gamma Z}_W$ are all loop functions \cite{FA1,FA2,FA3}. Based on the running quark mass and the low-energy values given in Ref.\cite{pdg2022}, the authors detailedly deduce the numerical values of $C^{SM}_{\gamma Z}$ and $\tilde{C}^{SM}_{\gamma Z}$ \cite{IN8}: $C^{SM}_{\gamma Z}{\sim}-2.395+0.001i$, $\tilde{C}^{SM}_{\gamma Z}{\sim}~0$.

\begin{figure}[ht]
\setlength{\unitlength}{5.0mm}
\centering
\includegraphics[width=5.0in]{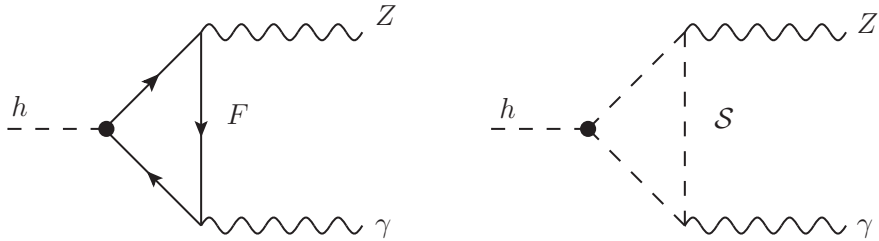}
\caption{The one-loop diagrams with new particles for the decay $ h\rightarrow Z \gamma$.}\label{N2}
\end{figure}

The Fig.\ref {N2} shows the NP one-loop diagrams of decay process $ h\rightarrow Z \gamma$ in the $U(1)_X$SSM. The charged fermion is expressed by $F$, and the charged scalar is denoted by $\mathcal{S}$. The new contributions to $C_{\gamma Z}$ come from the exchanged particles, including charginos, sleptons, squarks and charged Higgs. As mentioned in Ref.\cite{tt5}, the QCD corrections to the process $ h\rightarrow Z \gamma$ are about $0.1\%$, which is very small and can be safely ignored.

Actually, the extended gauge structure modifies the contribution of loops with the SM fermions to these couplings. Here, we show an example to see the modified effects.
The specific forms of $Z-\bar{u}_i-u_i$ in the SM and $U(1)_X$SSM are respectively,
\begin{eqnarray}
&&\mathcal{L}^{SM}_{Z\bar{u}u}=\bar{u}_i\Big[-\frac{ e}{2 \sin\theta_W \cos\theta_W} (1-\frac{4}{3}{\sin^2\theta_W}){\gamma}_{\mu}P_L+\frac{2e \sin\theta_W}{3\cos\theta_W}{\gamma}_{\mu}P_R \Big]u_iZ^\mu,\label{s1}
\\&&\mathcal{L}^{U(1)_X}_{Z\bar{u}u}=\frac{1}{6}\bar{u}_i\Big[(-3 g_2\cos\theta_W\cos\theta_W^\prime+g_1\sin\theta_W\cos\theta_W^\prime
-g_{YX}\sin\theta_W^\prime){\gamma}_{\mu}P_L \nonumber\\&&\hspace{1.3cm}+ \Big(-(4g_{YX}+3g_X)\sin\theta_W^\prime+4g_1\sin\theta_W\cos\theta_W^\prime\Big){\gamma}_{\mu}P_R \Big]u_i Z^{\mu}.\label{ss}
\end{eqnarray}

We can see that $\mathcal{L}^{U(1)_X}_{Z\bar{u}u}$ is related to $\theta_{W}'$, $g_X$ and $g_{YX}$, where $\theta_{W}'$ is the core parameter. $\theta_{W}'$ is the new mixing angle appearing in the couplings of $Z$ and $Z'$.
Supposing $\theta_{W}' = 0$, from Eq.(\ref{ss}) we can obtain
\begin{eqnarray}
&&\mathcal{L}^{U(1)_X}_{Z\bar{u}u}\rightarrow\frac{1}{6}\bar{u}_i\Big[(-3 g_2\cos\theta_W+g_1\sin\theta_W){\gamma}_{\mu}P_L+4g_1\sin\theta_W{\gamma}_{\mu}P_R \Big]u_i Z^{\mu}
\nonumber\\&&\hspace{1.3cm}=\bar{u}_i\Big[-\frac{ e}{2 \sin\theta_W \cos\theta_W} (1-\frac{4}{3}{\sin^2\theta_W}){\gamma}_{\mu}P_L+\frac{2e \sin\theta_W}{3\cos\theta_W}{\gamma}_{\mu}P_R \Big]u_i Z^{\mu}.\label{kk}
\end{eqnarray}

Obviously, Eq.(\ref{kk}) is the same as Eq.(\ref{s1}). The above analysis shows that $\mathcal{L}^{SM}_{Z\bar{u}u}$ and $\mathcal{L}^{U(1)_X}_{Z\bar{u}u}$
 are equal as $\theta_{W}'=0$. According to our numerical calculation, $\theta_{W}' \sim 10^{-5}$ in the $U(1)_X$SSM.
 That is to say, the ratio of the difference between $\mathcal{L}^{SM}_{Z\bar{u}u}$ and $\mathcal{L}^{U(1)_X}_{Z\bar{u}u}$ to $\mathcal{L}^{SM}_{Z\bar{u}u}$ $(\frac{\mathcal{L}^{SM}_{Z\bar{u}u}-\mathcal{L}^{U(1)_X}_{Z\bar{u}u}}{\mathcal{L}^{SM}_{Z\bar{u}u}})$ is at the order of $10^{-5}$. Therefore, the difference in this condition can be neglected safely.
  In the end, for the contributions of loops with the SM fermions to these couplings modified by the extended gauge structure, we adopt the results of the SM.

In the $U(1)_X$SSM, the expression of CP-even coupling $C^{U(1)_X}_{\gamma Z}$ is as follows
\begin{eqnarray}
&&\hspace{-0.3cm}C^{U(1)_X}_{\gamma Z}=\frac{v \sin\theta_W \cos\theta_W}{e}\int_{0}^{1}dx\int_{0}^{1}ydy \Big\{ \sum_{F={\chi}^{\pm}}\Big[ \frac{Q_{F1}}{R_{1N}^2(m_{F_1},m_{F_2})}\Big(A^{\bar{F}_2 F_1 h} B^{\bar{F}_1 F_2 Z}(-2(x-1)y^2\nonumber\\&&\hspace{1.0cm}(m_{F_1}+m_{F_2})+y(2 x(m_{F_1}+m_{F_2})-3 m_{F_1}-m_{F_2})+m_{F_1})+A_w^{\bar{F}_2 F_1 h} B_w^{\bar{F}_1 F_2 Z}(m_{F_1}(y-1)\nonumber\\&&\hspace{1.0cm}(2(x-1)y+1)+ym_{F_2}(-2xy+2x+2y-1))\Big)+\frac{Q_{F1}}{R_{2N}^2(m_{F_1},m_{F_2})}\Big(A^{\bar{F}_2 F_1 h} B^{\bar{F}_1 F_2 Z}(-2\nonumber\\&&\hspace{1.0cm}(x-1)y^2(m_{F_1}+m_{F_2})+y(x-1)(3m_{F_1}+m_{F_2})+m_{F_1})+A_w^{\bar{F}_2 F_1 h} B_w^{\bar{F}_1 F_2 Z}(m_{F_1}(y-1)\nonumber\\&&\hspace{1.0cm}(2(x-1)y+1)-y(m_{F_1}x+m_{F_2}(x-1)(2y-1)))\Big)\Big] +\sum_{\mathcal{S}=\tilde{L},\tilde{D},\tilde{U},H^{\pm}} Q_{S1} A^{S^*_2 S_1 h} B^{S^*_1 S_2 Z} \nonumber\\&&\hspace{1.0cm} \Big(\frac{1}{R_{1N}^2(m_{S_1},m_{S_2})}+\frac{1}{R_{2N}^2(m_{S_1},m_{S_2})}\Big)(1-x)y(1-y)\Big\}.
\end{eqnarray}

In the SM, the CP-odd coupling $\tilde{C}^{SM}_{\gamma Z}=0$. In other words, $\tilde{C}_{\gamma Z}$ only contains the new physics contributions $\tilde{C}^{U(1)_X}_{\gamma Z}$. In the $U(1)_X$SSM, the scalar loop represented by the right diagram in Fig.2 does not contribute to the CP-odd coupling. Therefore, among these new particles, only charginos provide corrections to $\tilde{C}^{U(1)_X}_{\gamma Z}$ in the one-loop diagrams.
\begin{eqnarray}
&&\hspace{-0.3cm}\tilde{C}^{U(1)_X}_{\gamma Z}=-\frac{i v \sin\theta_W \cos\theta_W}{e}\int_{0}^{1}dx \int_{0}^{1}ydy \sum_{F={\chi}^{\pm}}\Big[\frac{1}{R_{1N}^2(m_{F_1},m_{F_2})} \Big(A^{\bar{F}_2 F_1 h} B_w^{\bar{F}_1 F_2 Z}(y(m_{F_1}+m_{F_2})\nonumber\\&&\hspace{1.0cm}-m_{F_1})+A_w^{\bar{F}_2 F_1 h} B^{\bar{F}_1 F_2 Z}(m_{F_1}(1-y)+m_{F_2}y)\Big)+\frac{1}{R_{2N}^2(m_{F_1},m_{F_2})} \Big(A^{\bar{F}_2 F_1 h} B_w^{\bar{F}_1 F_2 Z}\nonumber\\&&\hspace{1.0cm}(y(1-x)(m_{F_1}+m_{F_2})-m_{F_1})+A_w^{\bar{F}_2 F_1 h} B^{\bar{F}_1 F_2 Z}((x-1)y(m_{F_1}-m_{F_2})+m_{F_1})\Big)\Big].
\end{eqnarray}

The functions $R_{1N}^2(m_{1},m_{2})$ and $R_{2N}^2(m_{1},m_{2})$ are shown as
\begin{eqnarray}
&&R_{1N}^2(m_{1},m_{2})=p^2_2(1-x)^2 y^2+p^2_1(1-y)^2-2 p_1{\cdot}p_2(1-x)y(1-y)+m^2_2xy\nonumber\\&&\hspace{2.8cm}+(m^2_2-p^2_2)(1-x)y+(m^2_1-p^2_1)(1-y),
\nonumber\\&&R_{2N}^2(m_{1},m_{2})=p^2_1(1-x)^2 y^2+p^2_2(1-y)^2-2 p_1{\cdot}p_2(1-x)y(1-y)+m^2_2xy\nonumber\\&&\hspace{2.8cm}+(m^2_2-p^2_1)(1-x)y+(m^2_1-p^2_2)(1-y).
\end{eqnarray}

The mass$(M)$ of particles in the loop is much larger than that of Higgs boson, and the mass ratio $\frac{m_h}{M} \sim 0.1$.
The momenta $(p_1, p_2)$ satisfy the on-shell condition $p^2_1=m^2_Z$ and $p^2_2=m^2_V$. From the relation
$p^2_2<p^2_1<m^2_h$, it is easy to find that the
particles in loops are sufficiently heavy and $p_1, p_2$ can be
considered as small parameters. In our calculation, we keep $p_1$ and $p_2$ terms with nonzero values
to obtain relatively accurate results.

The coupling constants of vertex $\bar{F}_2 F_1 h$ are $A^{\bar{F}_2 F_1 h}$ and $A_w^{\bar{F}_2 F_1 h}$. The coupling constants of vertex $\bar{F}_1 F_2 Z$ are $B^{\bar{F}_1 F_2 Z}$ and $B_w^{\bar{F}_1 F_2 Z}$. Their general forms are represented by
\begin{eqnarray}
\bar{F}_2 i (A^{\bar{F}_2 F_1 h}+A_w^{\bar{F}_2 F_1 h} {\gamma}_{5}) F_1 h,~~~
 \bar{F}_1 i(B^{\bar{F}_1 F_2 Z} {\gamma}_{\mu}+B_w^{\bar{F}_1 F_2 Z} {\gamma}_{\mu} {\gamma}_{5})F_2 Z^{\mu}.
\end{eqnarray}
We can find the required coupling vertexes in Sec.II and Appendix.A.

As discussed in Refs.\cite{IN8,sun4}, unlike the indirect contributions, the direct contributions of decay can be calculated in the power series of $(m_q/m_h)^2$ or $({\Lambda}_{QCD}/m_h)^2$. If the vector meson in the final state is longitudinally polarized, the direct contributions are produced from subleading-twist projections leading to suppressed power. For the transversely polarized vector meson, leading-twist projections provide direct contributions. We can get the concrete expression of direct contributions through the use of asymptotic function ${\phi}^{\perp}_V(x)=6x(1-x)$ \cite{FA4,FA5,FA6},

\begin{eqnarray}
&&F^{VZ}_{\perp direct}= \sum_{q} f^{q \perp}_V v_q {\kappa}_q \frac{3 m_q}{2 m_V} \frac{1-r^2_Z+2 r_Z \ln{r_Z}}{(1-r_Z)^2},
\nonumber\\&&\tilde{F}^{VZ}_{\perp direct}= \sum_{q} f^{q \perp}_V v_q \tilde{\kappa}_q \frac{3 m_q}{2 m_V} \frac{1-r^2_Z+2 r_Z \ln{r_Z}}{(1-r_Z)^2},
\end{eqnarray}
here, $m_q$ is the constituent quark mass in the meson, ${\Lambda}_{QCD}$ is the hadronic scale and $f^{q \perp}_V$ represents the flavor-specific transverse decay constants of the meson. This kind of direct contributions seem comparable with the indirect contributions in Eq.(\ref{f5}). Numerically, the direct contributions are still severely suppressed. Compared with indirect contributions, direct contributions are very small. Therefore, indirect contributions are more important than direct contributions.

In the last two diagrams of Fig.1, the $hZZ$ vertex exists at tree level. There is no $h \gamma Z$ coupling at tree level, but it can be produced by loop diagrams \cite{IN3,IN4}. So the Fig.1(d) should be subdominant compared to the Fig.1(c) on the level of magnitude.
Although the last diagram is subdominant, it has a significant impact, because the possible
new-physics (chargino, slepton, squark) contributions come from the crossed circle.

\section{Numerical analysis}

In this part, considering the following experimental constraints, we adopt the lightest CP-even Higgs mass $m_{h}$=125.1 GeV \cite{sun1,sun2}. The experimental value of $\tan{\beta}_{\eta}$ should be less than 1.5 \cite{TanBP}. According to the latest LHC data \cite{wx1,wx2,wx3,wx4,wx5,wx6,wx7}, we hold that the slepton mass is greater than $700~{\rm GeV}$, the chargino mass is greater than $1100 ~{\rm GeV}$, and the squark mass is greater than $1200~{\rm GeV}$. Combined with the above experimental requirements, we get abundant data and process the data to get interesting one-dimensional graphs and multidimensional scatter plots.

Next, we will discuss the numerical analysis in four parts: 1. determining the needed parameters in the research process; 2. discussing the decay processes $h\rightarrow \gamma \gamma$ and $h\rightarrow V V^*(V=Z,W)$; 3. discussing the decay process $h\rightarrow Z \gamma$; 4. discussing the decay processes $h\rightarrow m_V Z$ with $m_V$ denoting $\omega,\rho,\phi,J/{\psi}$ and $\Upsilon$.

\subsection{ The input parameters scheme }

We have fixed the used parameters in the $U(1)_X$SSM:
\begin{table}[ht]
\caption{ The vector mesons decay constants in the study}
\begin{tabular}{|c|c|c|c|c|c|}
\hline
Vector meson & $\omega$ & $\rho$ & $\phi$ & $J/\psi$ & $\Upsilon$ \\
\hline
$m_V/{\rm GeV}$ & 0.782 & 0.77 & 1.02 & 3.097 & 9.46 \\
\hline
$f_V/{\rm GeV}$ & 0.194 & 0.216 & 0.223 & 0.403 & 0.684 \\
\hline
$v_V$ &$ -\frac{\sin^2\theta_W}{3 \sqrt{2}}$ & $\frac{1}{\sqrt{2}} (\frac{1}{2}-\sin^2\theta_W)$ & $-\frac{1}{4}+\frac{\sin^2\theta_W}{3}$ & $\frac{1}{4}-\frac{2 \sin^2\theta_W}{3}$ & $-\frac{1}{4}+\frac{\sin^2\theta_W}{3}$  \\
\hline
$Q_V$ & $\frac{1}{3 \sqrt{2}}$ & $\frac{1}{\sqrt{2}}$ & $-\frac{1}{3}$ & $\frac{2}{3}$ & $-\frac{1}{3}$ \\
\hline
${f^{\perp}_V}/{f_V}={f^{q \perp}_V}/{f^{q}_V}$ & 0.71 & 0.72 & 0.76 & 0.91 & 1.09 \\
\hline
\end{tabular}
\label{t1}
\end{table}
\begin{eqnarray}
&&\lambda_H = -0.1,~\lambda_C = -0.25,~v_S=5.5~{\rm TeV},~\kappa=0.1,~T_{\lambda_H} = 1.5~{\rm TeV},~T_{\lambda_C} = 1~{\rm TeV},
\nonumber\\&&l_W= 10~{\rm TeV}^2,~ B_{\mu} = 2~{\rm TeV}^2,~B_S=0.1~{\rm TeV}^2,~M_1 =0.4~{\rm TeV},~ M_{BL}=0.4~{\rm TeV},
\nonumber\\&&M_{BB'}=0.2~{\rm TeV},
~{\mu}=1.5~{\rm TeV},~M_S =1 ~{\rm TeV},~T_{\kappa} =9~ {\rm TeV},~Y_{Xii}=1,
\nonumber\\&&M^2_{Qii}=3.8~{\rm TeV}^2,~M^2_{Uii}=16~{\rm TeV}^2,~M^2_{Dii}=1~{\rm TeV}^2,~M^2_{\tilde{E}ii}=7~{\rm TeV}^2,
\nonumber\\&&M^2_{{\tilde{\nu}}ii}=0.5~{\rm TeV}^2,~T_{uii}=32~{\rm TeV},~T_{{\nu}ii}=1~{\rm TeV},~T_{Xii}= -1~{\rm TeV},~(i=1,2,3).
\end{eqnarray}

Here, it is worth noting that all parameters' non-diagonal elements are assumed to be zero. We employ the following parameters as variable parameters in numerical analysis.
\begin{eqnarray}
&&\tan{\beta}_{\eta},~\tan{\beta},~g_{YX},~g_X,~M_2,~M^2_{\tilde{L}ii}=m^2_{\tilde{L}},~T_{eii}=T_e,~T_{dii}=T_d,~(i=1,2,3).
\end{eqnarray}

\subsection{ The decay processes $h\rightarrow \gamma \gamma$ and $h\rightarrow V V^*$ }

\begin{figure}[ht]
\setlength{\unitlength}{5mm}
\centering
\includegraphics[width=3.0in]{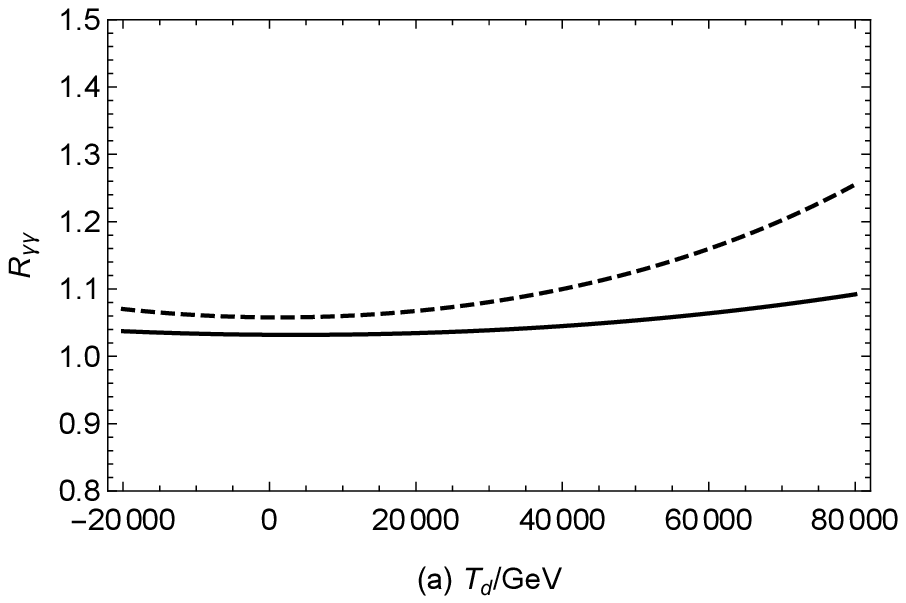}
\vspace{0.2cm}
\setlength{\unitlength}{5mm}
\centering
\includegraphics[width=3.0in]{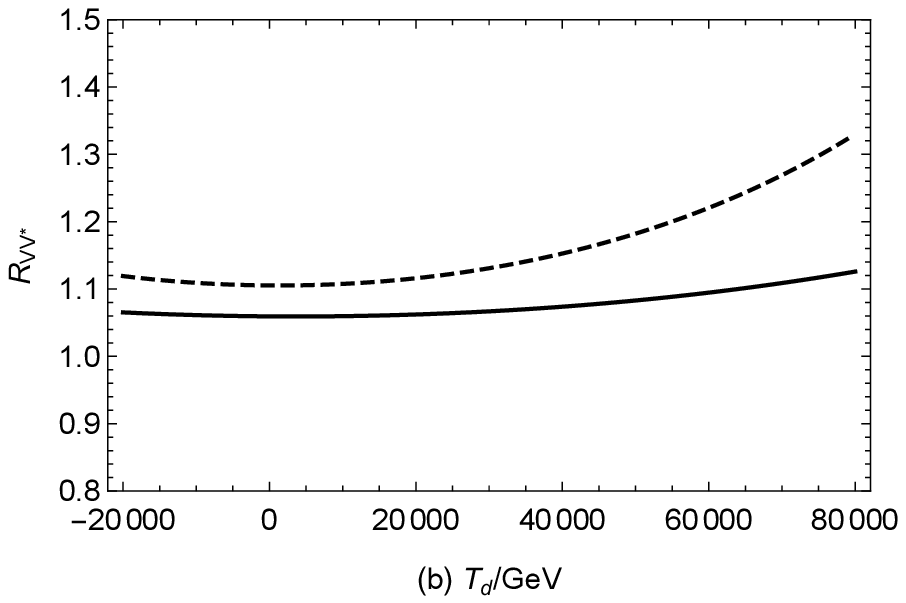}
\caption{(a) displays the dashed line ($\tan\beta=12$) and solid line ($\tan\beta=20$) in ($T_d,R_{\gamma \gamma}$) plane. (b) displays the dashed line ($\tan\beta=12$) and solid line ($\tan\beta=20$) in ($T_d,R_{V V^*}$) plane.}{\label {0}}
\end{figure}

In this subsection, we calculate the ratios $R_{\gamma\gamma}$ and $R_{V V^*}$ for the processes
$h\rightarrow \gamma \gamma$ and $h\rightarrow V V^*(V=Z,W)$, respectively. Some relevant formulas can be found in the works \cite{rr1,rr2,cao}. With the parameters $g_X=0.35,~g_{YX}=0.1,~\tan{\beta}_{\eta}=1.05,~T_e=12~{\rm TeV},~M_2=1~{\rm TeV},~m^2_{\tilde{L}}=8.2{\times}10^5~{\rm GeV}^2$, we paint $R_{\gamma \gamma}$ and $R_{V V^*}$ versus $T_d$ in Fig.\ref {0}. The dashed and solid lines correspond to $\tan\beta=12$ and $\tan\beta=20$ in the diagram.

In Fig.\ref {0}(a), both curves are above $R_{\gamma \gamma}=1.03$. The solid line and the dashed line
are of similar behavior versus $T_d$, and they are very near in $-20000~{\rm GeV}<T_d<20000~{\rm GeV}$ area. When $T_d>20000~{\rm GeV}$, there is a clear upward trend in both solid and dashed lines, and the distance between the two lines gradually increases. The dashed line varies from 1.07 to 1.26 and the solid line varies from 1.04 to 1.10.

In Fig.\ref {0}(b), both curves are above $R_{V V^*}=1.05$. It is obvious that both the solid line and the dashed line turn large mightily in the $T_d$ region of $(20000~{\rm GeV},80000~{\rm GeV})$. The dashed line grows faster than the solid line. The biggest and smallest values of the dashed line are 1.34 and 1.12, respectively. The values of the solid line vary from 1.07 to 1.13. On the whole, our results for the processes
$h\rightarrow \gamma \gamma$ and $h\rightarrow V V^*(V=Z,W)$ can satisfy their experimental constraints.

\subsection{ The decay process $h\rightarrow Z \gamma$ }

In the numerical calculation of the process $h\rightarrow Z \gamma$, we use the parameters as $M_2\!=\!1~{\rm TeV},~g_X\!=\!0.3,~g_{YX}\!=\!0.1,~T_e\!=\!12~{\rm TeV},~T_d=15~{\rm TeV}$ in Fig.\ref {1}. $\tan{\beta}_{\eta}$ is a crucial parameter that can affect the quality of particles by directly affecting $v_\eta$ and $v_{\bar{\eta}}$. Taking $\tan{\beta}_{\eta}$ in ($0.8 - 1.3$) can get more reasonable numerical results. With the parameters $m^2_{\tilde{L}}=8.2{\times}10^5~{\rm GeV}^2$, we plot {${\Gamma}_{U(1)_X}(h\rightarrow Z \gamma)/{\Gamma}_{SM}(h\rightarrow Z \gamma)$ versus $\tan{\beta}_{\eta}$ in the Fig.\ref {1}(a). The solid line corresponds to $\tan\beta=20 $ and the dashed line corresponds to $\tan\beta=40$. We can see that the two lines are basically coincident in $1.04<\tan{\beta}_{\eta}<1.1$ and increase significantly with the increasing $\tan{\beta}_{\eta}$ in the range of ($1.10-1.13$). The dashed curve is larger than the solid curve. The solid line can reach 1.2, and the dashed line can almost reach 1.3.

Then we analyze the effects of the parameter $m^2_{\tilde{L}}$ on ${\Gamma}_{U(1)_X}(h\rightarrow Z \gamma)/{\Gamma}_{SM}(h\rightarrow Z \gamma)$. $m^2_{\tilde{L}}$ has an influence on the masses of scalar leptons. Based on $\tan{\beta}_{\eta}=1.05$, the numerical results are shown in Fig.\ref {1}(b) by the solid curve and dashed curve corresponding to $\tan\beta=20$ and $\tan\beta=30$ respectively. ${\Gamma}_{U(1)_X}(h\rightarrow Z \gamma)/{\Gamma}_{SM}(h\rightarrow Z \gamma)$ varies with $m^2_{\tilde{L}}$ in the range from $3.5{\times}10^5~{\rm GeV}^2$ to $9.0{\times}10^5~{\rm GeV}^2$. It can be clearly seen that both the solid line and the dashed line have an obvious downward trend in $3.5{\times}10^5~{\rm GeV}^2<m^2_{\tilde{L}}<6.0{\times}10^5~{\rm GeV}^2$ and start to overlap when $m^2_{\tilde{L}}=6.0{\times}10^5~{\rm GeV}^2$. The slope of the dashed line is greater than that of the solid line. The solid line arrives at 1.28, and the dashed line can almost reach 1.34.

In summary, when $\tan{\beta}_{\eta}>1.12$ and $m^2_{\tilde{L}}<3.8{\times}10^5~{\rm GeV}^2$, the ratio deviates more than $20\%$ from the predictions of the SM. When $\tan{\beta}_{\eta}$ increases, the masses of particles decrease. When $m^2_{\tilde{L}}$ decreases, the masses of scalar leptons decrease. Thus, we can conclude that large $\tan{\beta}_{\eta}$ and small $m^2_{\tilde{L}}$ lead to the maximal effects.

\begin{figure}[ht]
\setlength{\unitlength}{5mm}
\centering
\includegraphics[width=2.85in]{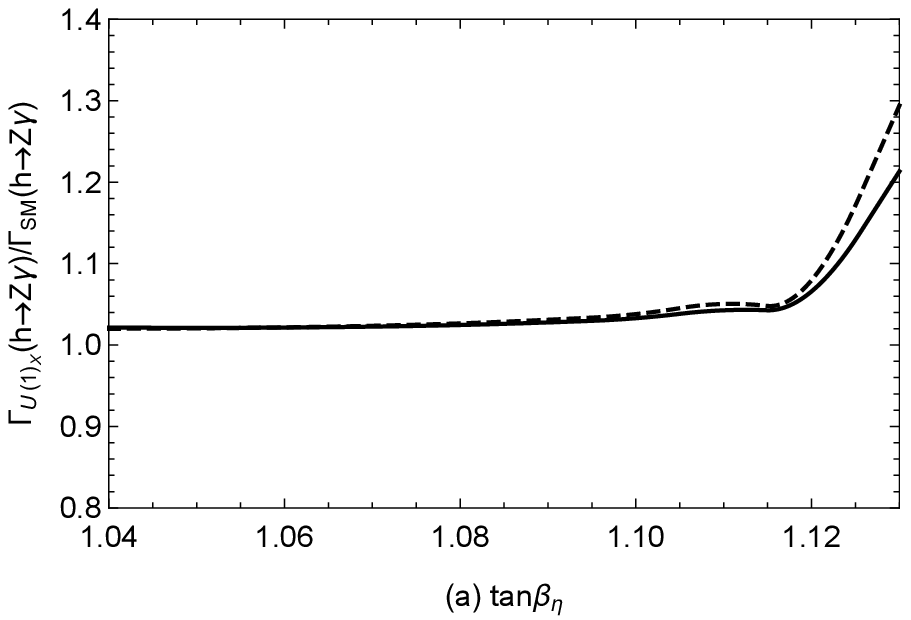}
\vspace{0.2cm}
\setlength{\unitlength}{5mm}
\centering
\includegraphics[width=3.0in]{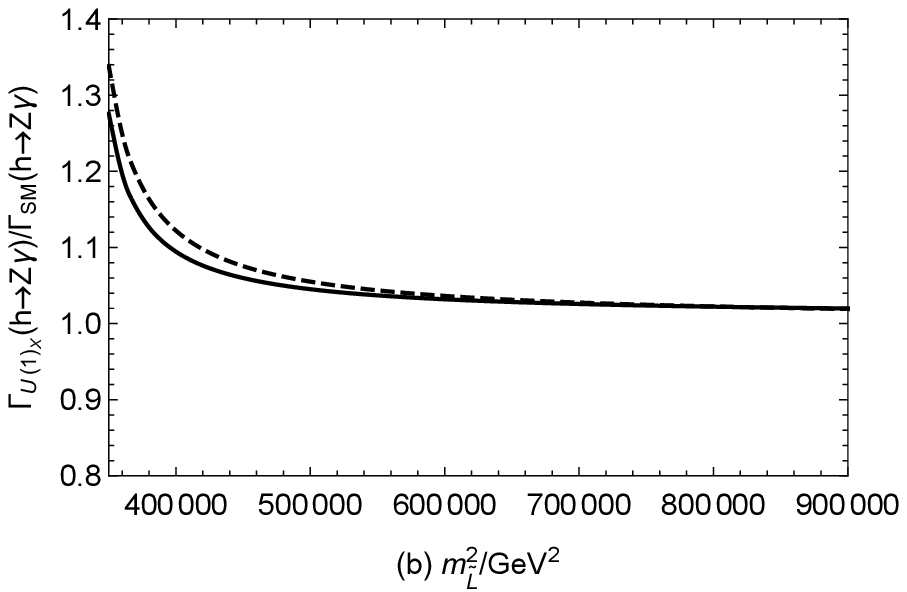}
\caption{(a) displays the solid line ($\tan\beta=20$) and dashed line ($\tan\beta=40$) in ($\tan{\beta}_{\eta},{\Gamma}_{U(1)_X}(h\rightarrow Z \gamma)/{\Gamma}_{SM}(h\rightarrow Z \gamma)$) plane. (b) displays the solid line ($\tan\beta=20$) and dashed line ($\tan\beta=30$) in ($m^2_{\tilde{L}},{\Gamma}_{U(1)_X}(h\rightarrow Z \gamma)/{\Gamma}_{SM}(h\rightarrow Z \gamma)$) plane.}{\label {1}}
\end{figure}

Next, we suppose the parameters with $M_2=1~{\rm TeV},~T_e=12~{\rm TeV},~T_d=15~{\rm TeV},~g_X=0.3$ and $\tan{\beta}=20$ in Fig.\ref{21}. We randomly scan some parameters, whose ranges are set as: $0.8<\tan{\beta}_{\eta}<1.3$, $0.05<g_{YX}<0.40$, and $3{\times}10^5~{\rm GeV}^2<m^2_{\tilde{L}}<5{\times}10^6~{\rm GeV}^2$. The meaning of \textcolor{blue}{$\blacksquare$} and \textcolor{orange}{$\bullet$} is given in Table \ref {t12}. With $g_{YX}=0.1$, we show the relationship between $\tan{\beta}_{\eta}$ and $m^2_{\tilde{L}}$ in Fig.\ref {21}(a). It is easy to see that \textcolor{blue}{$\blacksquare$} and \textcolor{orange}{$\bullet$} are obviously layered. When $\tan{\beta}_{\eta}<1.0$, the space is filled with blue squares \textcolor{blue}{$\blacksquare$}. Most of the orange circles \textcolor{orange}{$\bullet$} are distributed in the range of $\tan{\beta}_{\eta}~(1.0-1.08)$. The number of \textcolor{orange}{$\bullet$} in this part decreases slowly with the increase of $m^2_{\tilde{L}}$. Numerically, the ratio of decay width is mostly in the region of ($1.05-1.30$) for large values.

With $\tan{\beta}_{\eta}=1.05$, Fig.\ref {21}(b) reflects the results in the plane of $g_{YX}$ versus $m^2_{\tilde{L}}$. The boundary between \textcolor{blue}{$\blacksquare$} and \textcolor{orange}{$\bullet$} is apparent. The line connected by (0.15, $3{\times}10^5$) and (0.40, $3.3{\times}10^6$) divides the space into two parts, with \textcolor{blue}{$\blacksquare$} and \textcolor{orange}{$\bullet$} on the above and below of the line respectively. This obviously indicates that large $g_{YX}$ and small $m^2_{\tilde{L}}$ can get greater contribution. The large values are in the range of ($1.05-1.30$) for the ratio of decay width.

\begin{table}[ht]
\caption{ The meaning of shape style in Fig.\ref{21}}
\begin{tabular}{|c|c|c|}
\hline
Shape style&Fig.\ref{21}(a)&Fig.\ref{21}(b)\\
\hline
\textcolor{blue}{$\blacksquare$} & ${\Gamma}_{U(1)_X}(h\rightarrow Z \gamma )/{\Gamma}_{SM}(h\rightarrow Z \gamma)<1.01$ & ${\Gamma}_{U(1)_X}(h\rightarrow Z \gamma )/{\Gamma}_{SM}(h\rightarrow Z \gamma)<1.02$ \\
\hline
\textcolor{orange}{$\bullet$}&${\Gamma}_{U(1)_X}(h\rightarrow Z \gamma )/{\Gamma}_{SM}(h\rightarrow Z \gamma)~{\ge}~1.01$ & ${\Gamma}_{U(1)_X}(h\rightarrow Z \gamma )/{\Gamma}_{SM}(h\rightarrow Z \gamma)~{\ge}~1.02$ \\
\hline
\end{tabular}
\label{t12}
\end{table}

\begin{figure}[ht]
\setlength{\unitlength}{5mm}
\centering
\includegraphics[width=3.1in]{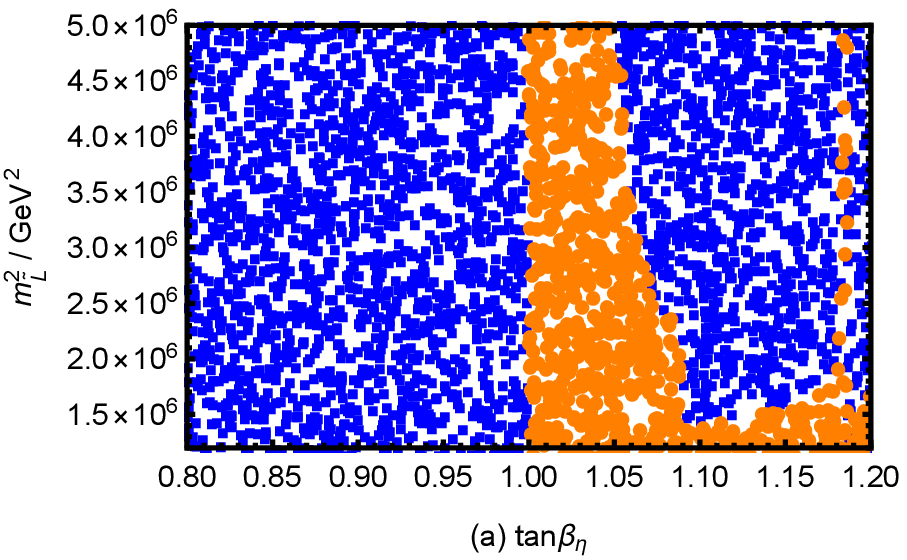}
\setlength{\unitlength}{5mm}
\centering
\includegraphics[width=3.1in]{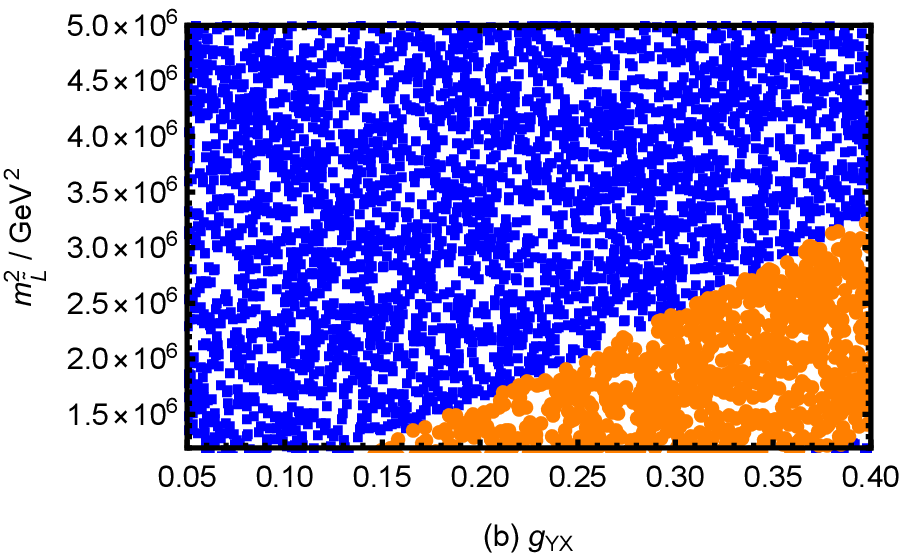}
\caption{${\Gamma}_{U(1)_X}(h\rightarrow Z \gamma)/{\Gamma}_{SM}(h\rightarrow Z \gamma)$ in $\tan{\beta}_{\eta}-m^2_{\tilde{L}}$ plane (a) and $g_{YX}-m^2_{\tilde{L}}$ plane (b).}{\label {21}}
\end{figure}

\subsection{ The decay processes of $h\rightarrow m_V Z$ }

We will analyze the decay processes $h\rightarrow m_V Z$ in this subsection. The vector mesons decay constants for $\omega,\rho,\phi,J/{\psi}$ and $\Upsilon$ can be found in Table \ref {t1}. At first, the process $h\rightarrow\omega Z$ is calculated. We use the parameters as $M_2=1~{\rm TeV},~T_e=12~{\rm TeV},~T_d=15~{\rm TeV},~m^2_{\tilde{L}}= 8.6{\times}10^5~{\rm GeV}^2$ and $\tan{\beta}_{\eta}=1.05$ in Fig.\ref {3}.

With $g_X=0.3$, we draw the solid line ($\tan\beta=20$) and dashed line ($\tan\beta=30$)  to represent the relationship between $g_{YX}$ and ${\Gamma}_{U(1)_X}(h\rightarrow \omega Z )/{\Gamma}_{SM}(h\rightarrow \omega Z)$ in Fig.\ref {3}(a). $g_{YX}$ is the coupling constant of gauge mixing that affects the strength of coupling vertexes. Both dashed line and solid line are in the  coincident state when $g_{YX}<0.22$. The two lines show an obvious upward trend in $0.20<g_{YX}<0.25$ area, and the increasing amplitude of dashed line is greater than that of solid line. Ultimately, the solid line and the dashed line can almost reach 1.43, which is due to the large $g_{YX}$.

Supposing $g_{YX}=0.1$, we plot $g_X$ versus ${\Gamma}_{U(1)_X}(h\rightarrow \omega Z )/{\Gamma}_{SM}(h\rightarrow \omega Z)$ in Fig.\ref {3}(b). The solid and dashed lines correspond to $\tan\beta=20$ and $\tan\beta=40$ in the right diagram. The solid line  coincides with the dashed line in $0.20<g_{X}<0.40$ area. The two lines turn big in the $g_X$ region of ($0.4-0.7$). The biggest and smallest values of the dashed line are 1.08 and 1.01  respectively. The values of the solid line vary from 1.01 to 1.03.

\begin{figure}[ht]
\setlength{\unitlength}{5mm}
\centering
\includegraphics[width=2.9in]{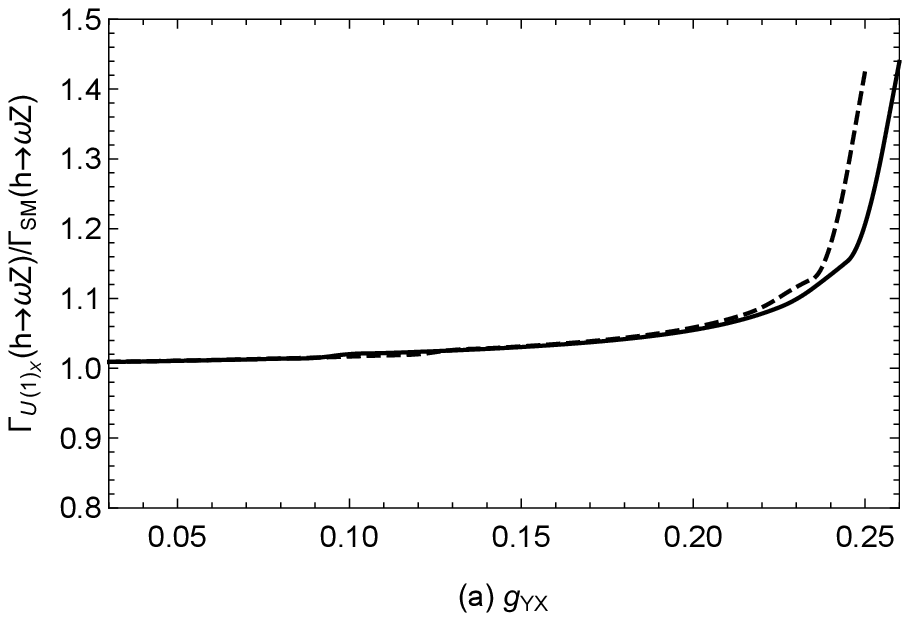}
\setlength{\unitlength}{5mm}
\centering
\includegraphics[width=3.0in]{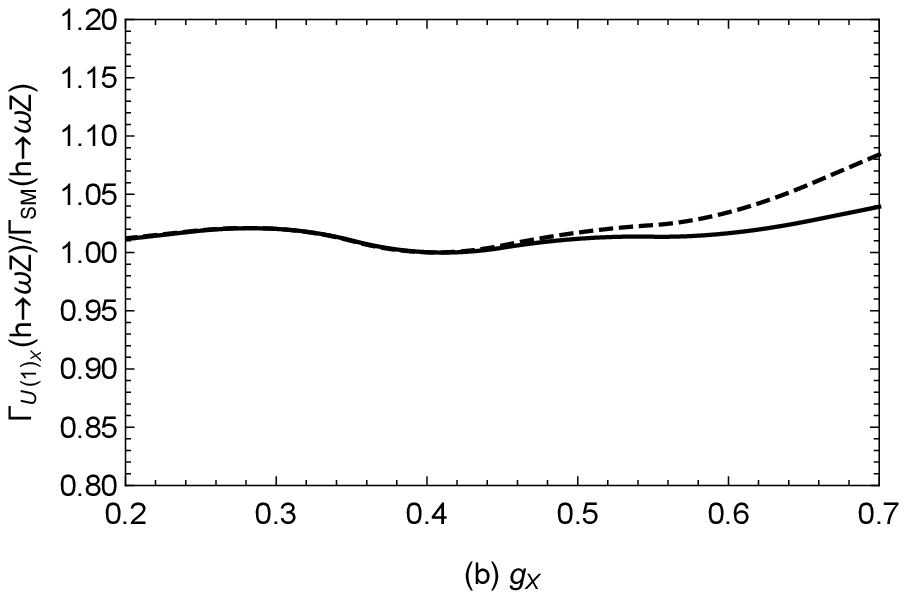}
\caption{(a) displays the solid line ($\tan\beta=20$) and dashed line ($\tan\beta=30$) in ($g_{YX},{\Gamma}_{U(1)_X}(h\rightarrow \omega Z )/{\Gamma}_{SM}(h\rightarrow \omega Z)$) plane. (b) displays the solid line ($\tan\beta=20$) and dashed line ($\tan\beta=40$) in ($g_X,{\Gamma}_{U(1)_X}(h\rightarrow \omega Z )/{\Gamma}_{SM}(h\rightarrow \omega Z)$) plane.}{\label {3}}
\end{figure}

We suppose the parameters with $M_2=1~{\rm TeV},~T_e=12~{\rm TeV},~T_d=15~{\rm TeV}$ and $\tan{\beta}=20$ in Fig.\ref{4}. To scan the parameter space better, some parameters ranges are set as: $0.8<\tan{\beta}_{\eta}<1.3$, $0.05<g_{YX}<0.40$, $0.3<g_{X}<0.7$, and $3{\times}10^5~{\rm GeV}^2<m^2_{\tilde{L}}<5{\times}10^6~{\rm GeV}^2$. The meaning of \textcolor{blue}{$\blacksquare$} and \textcolor{orange}{$\bullet$} can be found in Table \ref {t11}. With $g_{X}=0.3$ and $m^2_{\tilde{L}}= 8.6{\times}10^5~{\rm GeV}^2$, we show ${\Gamma}_{U(1)_X}(h\rightarrow \omega Z )/{\Gamma}_{SM}(h\rightarrow \omega Z)$ in the plane of $\tan{\beta}_{\eta}$ versus $g_{YX}$ in Fig.\ref{4}(a). The space is roughly divided into six parts, and \textcolor{orange}{$\bullet$} occupy more space than \textcolor{blue}{$\blacksquare$}. Based on the numerical value, we know that the ratio ranges roughly from 1.05 to 1.30. As $g_{YX}=0.1$ and $\tan{\beta}_{\eta}=1.05$, we draw ${\Gamma}_{U(1)_X}(h\rightarrow \omega Z )/{\Gamma}_{SM}(h\rightarrow \omega Z)$ in the plane of $g_{X}$ versus $m^2_{\tilde{L}}$ in Fig.\ref{4}(b). \textcolor{orange}{$\bullet$} are mainly concentrated in the area $g_{X}$ (0.46, 0.60) and $m^2_{\tilde{L}}~(3{\times}10^5,~2{\times}10^6)~{\rm GeV}^2$. \textcolor{blue}{$\blacksquare$} are located in the remaining space. The large values range from 1.05 to 1.30 for the ratio of decay width.
\begin{table}[ht]
\caption{ The meaning of shape style in Fig.\ref{4}}
\begin{tabular}{|c|c|c|}
\hline
Shape style&Fig.\ref{4}(a)&Fig.\ref{4}(b)\\
\hline
\textcolor{blue}{$\blacksquare$} & ${\Gamma}_{U(1)_X}(h\rightarrow \omega Z )/{\Gamma}_{SM}(h\rightarrow \omega Z)<1.0$ & ${\Gamma}_{U(1)_X}(h\rightarrow \omega Z )/{\Gamma}_{SM}(h\rightarrow \omega Z)<1.005$ \\
\hline
\textcolor{orange}{$\bullet$} &${\Gamma}_{U(1)_X}(h\rightarrow \omega Z )/{\Gamma}_{SM}(h\rightarrow \omega Z)~{\ge}~1.0$ & ${\Gamma}_{U(1)_X}(h\rightarrow \omega Z )/{\Gamma}_{SM}(h\rightarrow \omega Z)~{\ge}~1.005$ \\
\hline
\end{tabular}
\label{t11}
\end{table}
\begin{figure}[ht]
\setlength{\unitlength}{5mm}
\centering
\includegraphics[width=3.0in]{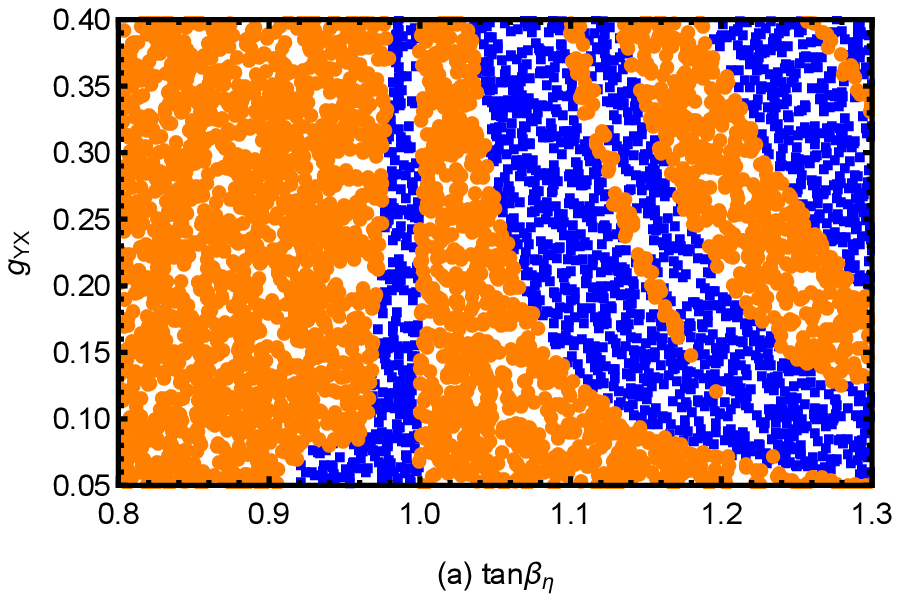}
\setlength{\unitlength}{5mm}
\centering
\includegraphics[width=3.2in]{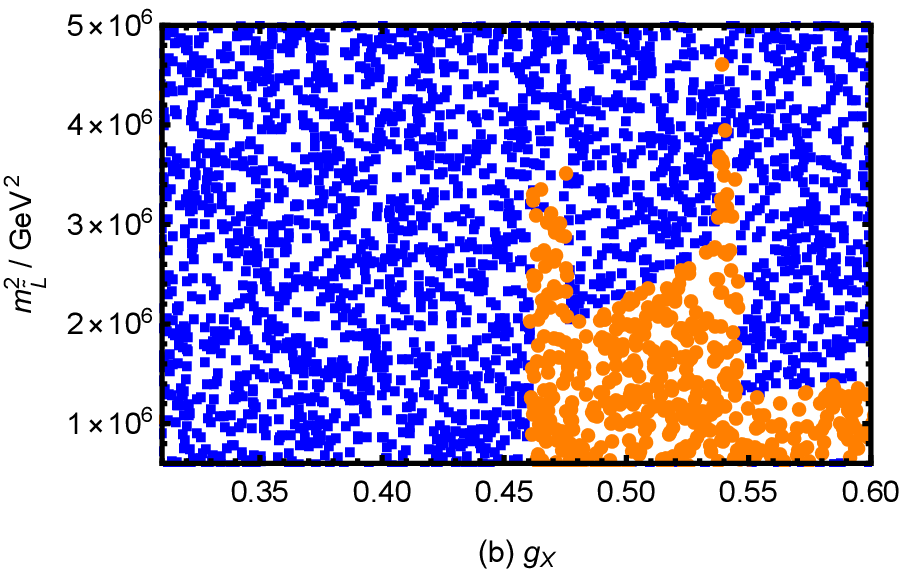}
\caption{${\Gamma}_{U(1)_X}(h\rightarrow \omega Z )/{\Gamma}_{SM}(h\rightarrow \omega Z)$ in $\tan{\beta}_{\eta}-g_{YX}$ plane (a) and $g_{X}-m^2_{\tilde{L}}$ plane (b).}{\label {4}}
\end{figure}

Secondly, we analyze the numerical results of the decay process $h\rightarrow \rho Z$. We suppose the parameters with $g_X=0.3,~g_{YX}=0.1,~T_e=12~{\rm TeV},~T_d=15~{\rm TeV}$ and $\tan{\beta}=20$ in Fig.\ref{5}. The scanned points are in the parameter spaces based on $0.8<\tan{\beta}_{\eta}<1.3$, $3{\times}10^5~{\rm GeV}^2<m^2_{\tilde{L}}<5{\times}10^6~{\rm GeV}^2$, and $300 ~{\rm GeV} <M_2< 3000 ~{\rm GeV}$. We can find the meaning of \textcolor{blue}{$\blacksquare$} and \textcolor{orange}{$\bullet$} in Table \ref {t2}. With $m^2_{\tilde{L}}=8.6{\times}10^5~{\rm GeV}^2$, the Fig.\ref{5}(a) shows contribution from $\tan{\beta}_{\eta}$ and $M_2$. This diagram is divided into four parts with clear boundaries. \textcolor{orange}{$\bullet$} concentrate in $\tan{\beta}_{\eta}$ (0.8, 0.96) and $\tan{\beta}_{\eta}$ (1.0, 1.14), respectively. \textcolor{blue}{$\blacksquare$} concentrate in the narrow area of $\tan{\beta}_{\eta}$ (0.96, 1.0) and $\tan{\beta}_{\eta}$ (1.14, 1.2). By contrast, \textcolor{orange}{$\bullet$} occupy more space. The ratio of decay width is mainly concentrated in the range from 1.05 to 1.40. As $\tan{\beta}_{\eta}=1.05$, we obtain ${\Gamma}_{U(1)_X}(h\rightarrow \rho Z )/{\Gamma}_{SM}(h\rightarrow \rho Z)$ in the plane of $m^2_{\tilde{L}}$ versus $M_2$ in Fig.\ref{5}(b).  \textcolor{blue}{$\blacksquare$} are distributed throughout the space. In contrast, \textcolor{blue}{$\blacksquare$} are denser in $3{\times}10^6~{\rm GeV}^2<m^2_{\tilde{L}}<5{\times}10^6~{\rm GeV}^2$. In $3{\times}10^5~{\rm GeV}^2<m^2_{\tilde{L}}<3{\times}10^6~{\rm GeV}^2$, \textcolor{orange}{$\bullet$} are more concentrated. When the value of $m^2_{\tilde{L}}$ increases, the number of \textcolor{orange}{$\bullet$} gradually decreases. Through the above numerical analysis, the ratio of decay width is mostly located in the range of $( 1.02-1.30 )$.

\begin{table}[ht]
\caption{ The meaning of shape style in Fig.\ref{5}}
\begin{tabular}{|c|c|c|}
\hline
Shape style&Fig.\ref{5}(a)&Fig.\ref{5}(b)\\
\hline
\textcolor{blue}{$\blacksquare$} & ${\Gamma}_{U(1)_X}(h\rightarrow \rho Z )/{\Gamma}_{SM}(h\rightarrow \rho Z)<1.0$ & ${\Gamma}_{U(1)_X}(h\rightarrow \rho Z )/{\Gamma}_{SM}(h\rightarrow \rho Z)<1.02$ \\
\hline
\textcolor{orange}{$\bullet$} &${\Gamma}_{U(1)_X}(h\rightarrow \rho Z )/{\Gamma}_{SM}(h\rightarrow \rho Z)~{\ge}~1.0$ & ${\Gamma}_{U(1)_X}(h\rightarrow \rho Z )/{\Gamma}_{SM}(h\rightarrow \rho Z)~{\ge}~1.02$ \\
\hline
\end{tabular}
\label{t2}
\end{table}

\begin{figure}[ht]
\setlength{\unitlength}{5mm}
\centering
\includegraphics[width=3.0in]{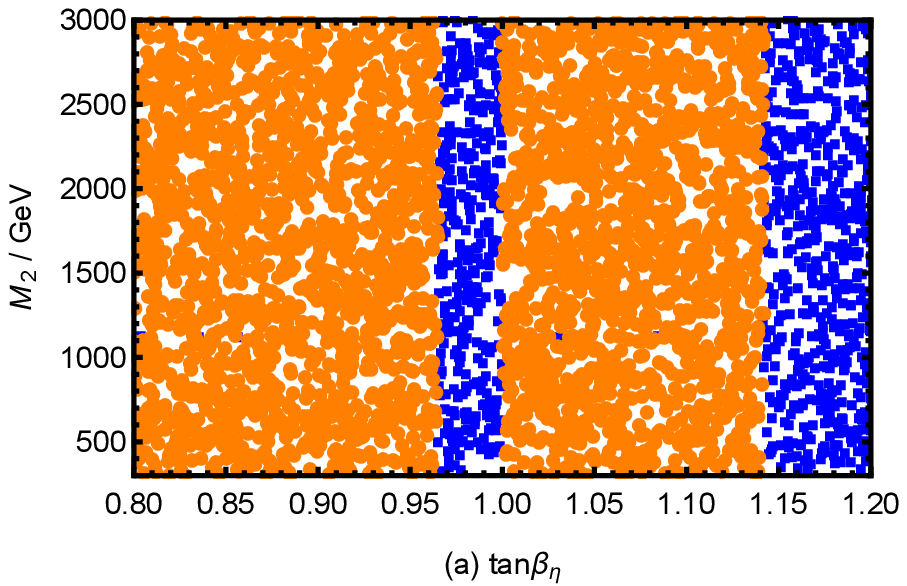}
\setlength{\unitlength}{5mm}
\centering
\includegraphics[width=3.05in]{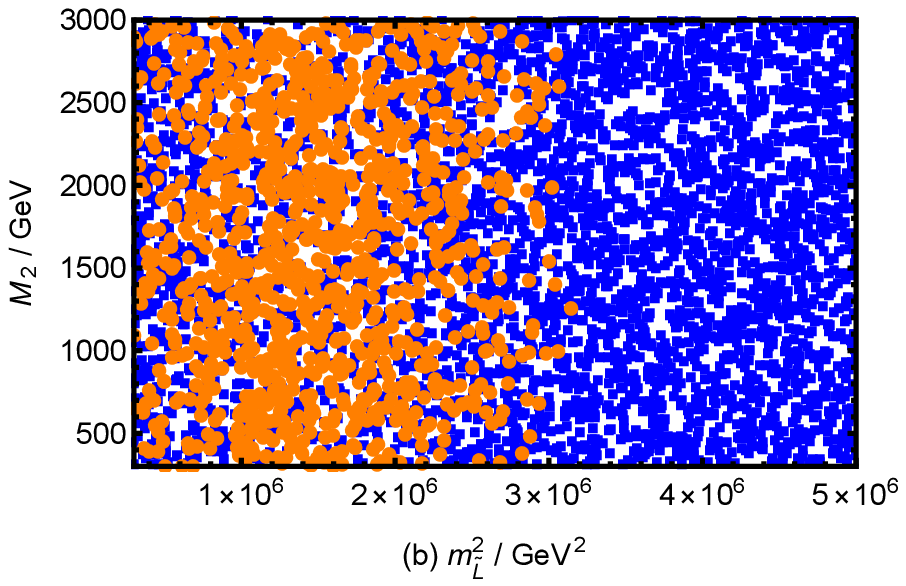}
\caption{${\Gamma}_{U(1)_X}(h\rightarrow \rho Z )/{\Gamma}_{SM}(h\rightarrow \rho Z)$ in $\tan{\beta}_{\eta}-M^2$ plane (a) and $m^2_{\tilde{L}}-M^2$ plane (b).}{\label {5}}
\end{figure}

Thirdly, the decay process $h\rightarrow\phi Z$ is summarized. We suppose the parameters with $M_2=1~{\rm TeV},~T_d=15~{\rm TeV},~\tan{\beta}_{\eta}=1.05$, and $\tan{\beta}=20$ in Fig.\ref{6}. The scanned parameters are: $0.3<g_X<0.7$,~$0.05<g_{YX}<0.40$,~$3{\times}10^5~{\rm GeV}^2<m^2_{\tilde{L}}<5{\times}10^6~{\rm GeV}^2$, and $-5000 ~{\rm GeV} <T_e< 5000 ~{\rm GeV}$. The meaning of \textcolor{green}{$\blacktriangle$}, \textcolor{blue}{$\blacksquare$} and \textcolor{orange}{$\bullet$} can be found in Table \ref {t3}. Supposing $g_X=0.3$ and $g_{YX}=0.1$, we plot $m^2_{\tilde{L}}$ varying with $T_e$ in Fig.\ref{6}(a). We can see that \textcolor{orange}{$\bullet$} are scattered in $3{\times}10^5~{\rm GeV}^2<m^2_{\tilde{L}}<8{\times}10^5~{\rm GeV}^2$, and \textcolor{blue}{$\blacksquare$} mainly concentrate in $8{\times}10^5~{\rm GeV}^2<m^2_{\tilde{L}}<4.2{\times}10^6~{\rm GeV}^2$ , and \textcolor{green}{$\blacktriangle$} are more concentrated in $4.2{\times}10^6~{\rm GeV}^2<m^2_{\tilde{L}}<5{\times}10^6~{\rm GeV}^2$. With the increase of $m^2_{\tilde{L}}$, ${\Gamma}_{U(1)_X}(h\rightarrow \phi Z )/{\Gamma}_{SM}(h\rightarrow \phi Z)$ decreases. These results imply that $m^2_{\tilde{L}}$ effect to ${\Gamma}_{U(1)_X}(h\rightarrow \phi Z )/{\Gamma}_{SM}(h\rightarrow \phi Z)$ is strong, but $T_e$ effect to ${\Gamma}_{U(1)_X}(h\rightarrow \phi Z )/{\Gamma}_{SM}(h\rightarrow \phi Z)$ is gentle. For the ratio of decay width, its middle values are in the range of ($1.01-1.02 $), and its large values are mainly distributed in the range of ($1.02 - 1.30 $). As $T_e=12~{\rm TeV}$, Fig.\ref {6}(b) shows  $m^2_{\tilde{L}}$ versus $g_X$, where \textcolor{orange}{$\bullet$} are concentrated in the upper left corner. The whole space is filled with \textcolor{blue}{$\blacksquare$}. This indicates that when the value of $m^2_{\tilde{L}}$ is smaller and the value of $g_X$ is bigger, its theoretical contribution can be improved. Our numerical results show that the large values range from 1.05 to 1.30.

\begin{table}[ht]
\caption{ The meaning of shape style in Fig.\ref{6}}
\begin{tabular}{|c|c|c|}
\hline
Shape style&Fig.\ref{6}(a)&Fig.\ref{6}(b)\\
\hline
\textcolor{green}{$\blacktriangle$} & ${\Gamma}_{U(1)_X}(h\rightarrow \phi Z )/{\Gamma}_{SM}(h\rightarrow \phi Z)<1.01$ & $\backslash$ \\
\hline
\textcolor{blue}{$\blacksquare$} & $1.01~{\le}~{\Gamma}_{U(1)_X}(h\rightarrow \phi Z )/{\Gamma}_{SM}(h\rightarrow \phi Z)<1.02$ & ${\Gamma}_{U(1)_X}(h\rightarrow \phi Z )/{\Gamma}_{SM}(h\rightarrow \phi Z)<1.02$ \\
\hline
\textcolor{orange}{$\bullet$} &${\Gamma}_{U(1)_X}(h\rightarrow \phi Z )/{\Gamma}_{SM}(h\rightarrow \phi Z)~{\ge}~1.02$ & ${\Gamma}_{U(1)_X}(h\rightarrow \phi Z )/{\Gamma}_{SM}(h\rightarrow \phi Z)~{\ge}~1.02$ \\
\hline
\end{tabular}
\label{t3}
\end{table}
\begin{figure}[ht]
\setlength{\unitlength}{5mm}
\centering
\includegraphics[width=3.08in]{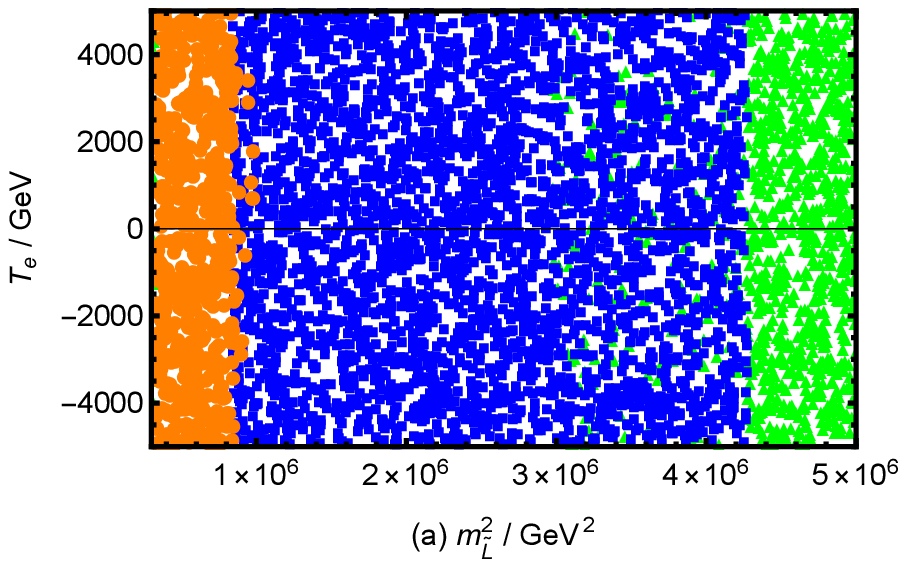}
\setlength{\unitlength}{5mm}
\centering
\includegraphics[width=3.0in]{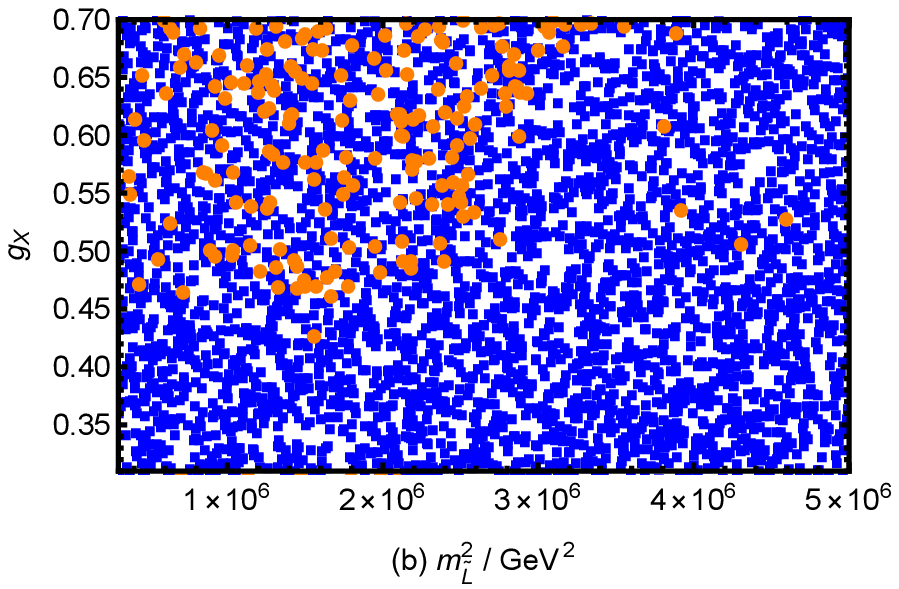}
\caption{${\Gamma}_{U(1)_X}(h\rightarrow \phi Z )/{\Gamma}_{SM}(h\rightarrow \phi Z)$ in $m^2_{\tilde{L}}-T_e$ plane (a) and $m^2_{\tilde{L}}-g_X$ plane (b).}{\label {6}}
\end{figure}

Then, we numerically analyze the decay process $h\rightarrow J/\psi Z$. The parameters are taken as $M_2=1~{\rm TeV},~g_X=0.1,~T_d=15~{\rm TeV}$, and $\tan{\beta}=20$ in Fig.\ref{7}. Some parameters ranges are set as: $0.8<\tan{\beta}_{\eta}<1.3$, $0.05<g_{YX}<0.40$, $-5000~{\rm GeV}<T_e<5000~{\rm GeV}$, and $3{\times}10^5~{\rm GeV}^2<m^2_{\tilde{L}}<5{\times}10^6~{\rm GeV}^2$. Table \ref{t4} indicates the meaning of \textcolor{green} {$\blacktriangle$}, \textcolor{blue}{$\blacksquare$} and \textcolor{orange}{$\bullet$}. Adopting $\tan{\beta}_{\eta}=1.05$ and $m^2_{\tilde{L}}=8.6{\times}10^5~{\rm GeV}^2$, we show the points in the plane of $g_{YX}$ and $T_e$ in Fig.\ref {7}(a). When $0.05<g_{YX}<0.2$, the space is filled with \textcolor{blue}{$\blacksquare$}.
In the range $0.2<g_{YX}<0.27$, \textcolor{orange}{$\bullet$} occupy much space.
\textcolor{green} {$\blacktriangle$} concentrate in the area $0.27<g_{YX}<0.4$. The boundary between \textcolor{green} {$\blacktriangle$}, \textcolor{blue}{$\blacksquare$} and \textcolor{orange}{$\bullet$} are obvious. The results imply that with the increase of $g_{YX}$, ${\Gamma}_{U(1)_X}(h\rightarrow J/\psi Z )/{\Gamma}_{SM}(h\rightarrow J/\psi Z)$ has a trend of first increasing and then decreasing. The middle values range from 1.0 to 1.05 and the larger values are from 1.05 to 1.45. With $g_{YX}=0.3$, we draw the points in the plane of $\tan{\beta}_{\eta}$ and $T_e$ in Fig.\ref {7}(b). \textcolor{green}{$\blacktriangle$} concentrate in the area $0.94<\tan{\beta}_{\eta}<1.0$, and \textcolor{blue}{$\blacksquare$} occupy in the remaining space. With the increase of $\tan{\beta}_{\eta}$,
for ${\Gamma}_{U(1)_X}(h\rightarrow J/\psi Z )/{\Gamma}_{SM}(h\rightarrow J/\psi Z)$, there is a trend of first decreasing and then increasing. The approximate range of the ratio is $(1.05-1.35)$.
\begin{table}[ht]
\caption{ The meaning of shape style in Fig.\ref{7}}
\resizebox{16.5cm}{!}{
\begin{tabular}{|c|c|c|}
\hline
Shape style&Fig.\ref{7}(a)&Fig.\ref{7}(b)\\
\hline
\textcolor{green}{$\blacktriangle$} & ${\Gamma}_{U(1)_X}(h\rightarrow J/\psi Z )/{\Gamma}_{SM}(h\rightarrow J/\psi Z)<1.0$ & ${\Gamma}_{U(1)_X}(h\rightarrow J/\psi Z )/{\Gamma}_{SM}(h\rightarrow J/\psi Z)<1.0$ \\
\hline
\textcolor{blue}{$\blacksquare$} & $1.0~{\le}~{\Gamma}_{U(1)_X}(h\rightarrow J/\psi Z )/{\Gamma}_{SM}(h\rightarrow J/\psi Z)<1.05$ & ${\Gamma}_{U(1)_X}(h\rightarrow J/\psi Z )/{\Gamma}_{SM}(h\rightarrow J/\psi Z)~{\ge}~1.0$ \\
\hline
\textcolor{orange}{$\bullet$} &${\Gamma}_{U(1)_X}(h\rightarrow J/\psi Z )/{\Gamma}_{SM}(h\rightarrow J/\psi Z)~{\ge}~1.05$ & $\backslash$ \\
\hline
\end{tabular}
\label{t4}}
\end{table}
\begin{figure}[ht]
\setlength{\unitlength}{5mm}
\centering
\includegraphics[width=3.05in]{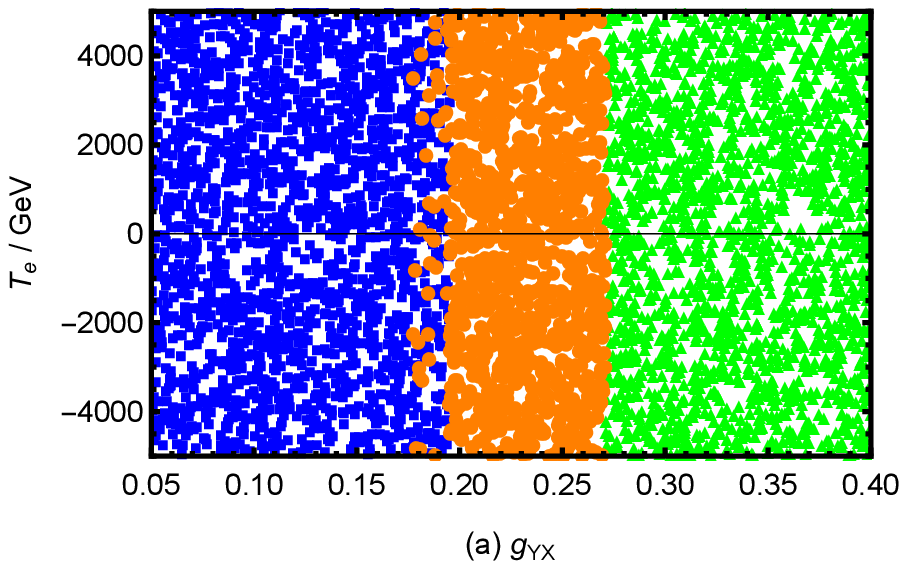}
\setlength{\unitlength}{5mm}
\centering
\includegraphics[width=3.0in]{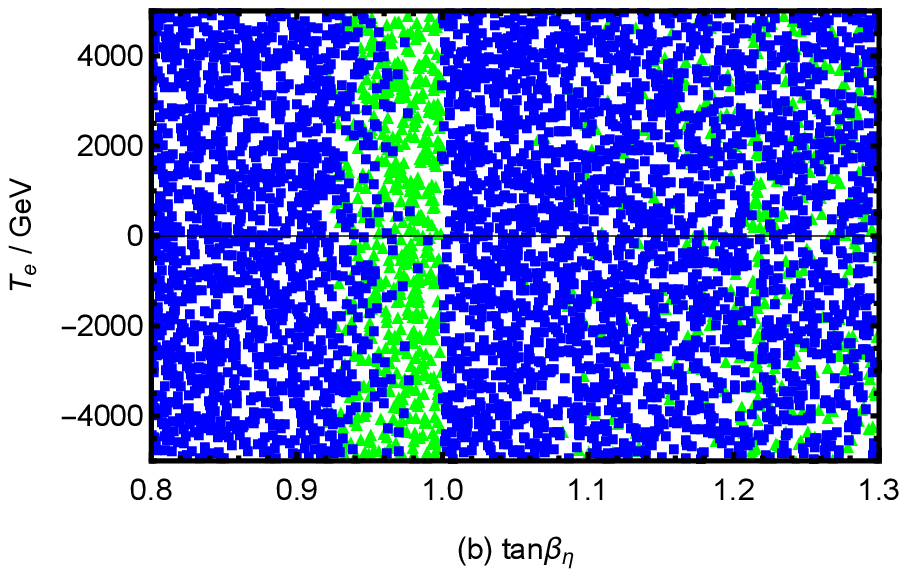}
\caption{${\Gamma}_{U(1)_X}(h\rightarrow J/\psi Z )/{\Gamma}_{SM}(h\rightarrow J/\psi Z)$ in $g_{YX}-T_e$ plane (a) and $\tan{\beta}_{\eta}-T_e$ plane (b).}{\label {7}}
\end{figure}

At last, we research the decay $h\rightarrow \Upsilon Z$ numerically. The parameters are taken as $M_2=1~{\rm TeV},~T_e=12~{\rm TeV},~T_d=15~{\rm TeV},~\tan{\beta}_{\eta}=1.05$, and $\tan{\beta}=20$ in Fig.\ref{8}. Some parameters ranges are set as: $0.05<g_{YX}<0.40$, $0.2<g_{X}<0.7$, and $3{\times}10^5~{\rm GeV}^2<m^2_{\tilde{L}}<5{\times}10^6~{\rm GeV}^2$. Table \ref{t5} gives the meaning of \textcolor{blue}{$\blacksquare$} and \textcolor{orange}{$\bullet$}. With $g_{YX}=0.1$, Fig.\ref{8}(a) shows $g_{X}$ versus $m^2_{\tilde{L}}$. \textcolor{orange}{$\bullet$} and \textcolor{blue}{$\blacksquare$} divide the space into two parts, and the former is slightly larger than the latter.
\textcolor{blue}{$\blacksquare$} are more concentrated in $0.3<g_{X}<0.46$, and \textcolor{orange}{$\bullet$} mainly concentrate in $0.46<g_{X}<0.7$. These suggest that with the increase of $g_{X}$, the ratio of decay width tends to increase. Based on the numerical value, the approximate range of the ratio is $(1.002-1.20)$. Adopting $m^2_{\tilde{L}}=8.6{\times}10^5~{\rm GeV}^2$, Fig.\ref{8}(b) shows $g_{YX}$ versus $g_{X}$. We can see that larger corrections (\textcolor{orange}{$\bullet$}) appear at the upper left corner. It shows that large $g_{X}$ and small $g_{YX}$ can improve the theoretical corrections. The larger value of the ratio of decay width ranges from 1.005 to 1.30.

\begin{table}[ht]
\caption{ The meaning of shape style in Fig.\ref{8}}
\begin{tabular}{|c|c|c|}
\hline
Shape style&Fig.\ref{8}(a)&Fig.\ref{8}(b)\\
\hline
\textcolor{blue}{$\blacksquare$} & ${\Gamma}_{U(1)_X}(h\rightarrow \Upsilon Z )/{\Gamma}_{SM}(h\rightarrow \Upsilon Z)<1.002$ & ${\Gamma}_{U(1)_X}(h\rightarrow \Upsilon Z )/{\Gamma}_{SM}(h\rightarrow \Upsilon Z)<1.005$ \\
\hline
\textcolor{orange}{$\bullet$} & ${\Gamma}_{U(1)_X}(h\rightarrow \Upsilon Z )/{\Gamma}_{SM}(h\rightarrow \Upsilon Z)~{\ge}~1.002 $& ${\Gamma}_{U(1)_X}(h\rightarrow \Upsilon Z )/{\Gamma}_{SM}(h\rightarrow \Upsilon Z)~{\ge}~1.005$ \\
\hline
\end{tabular}
\label{t5}
\end{table}
\begin{figure}[ht]
\setlength{\unitlength}{5mm}
\centering
\includegraphics[width=3.1in]{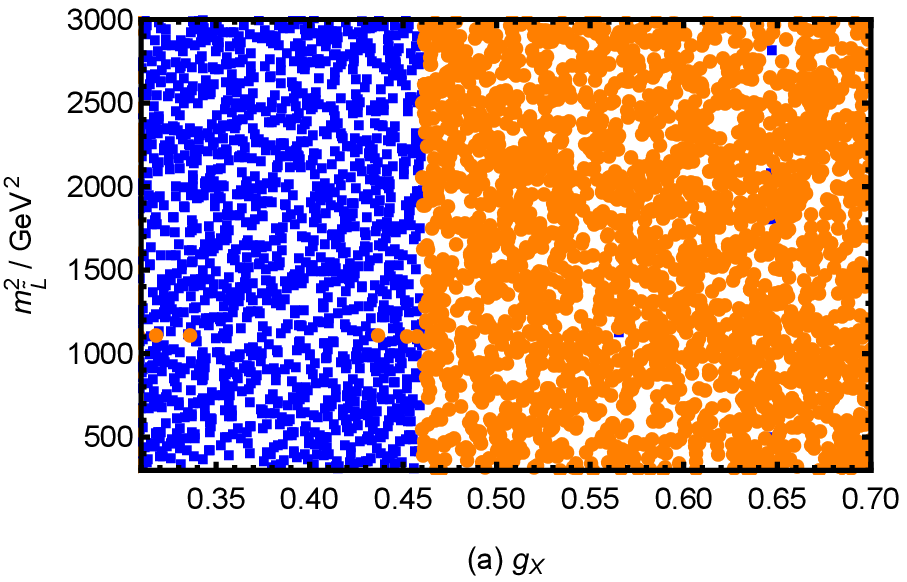}
\setlength{\unitlength}{5mm}
\centering
\includegraphics[width=3.0in]{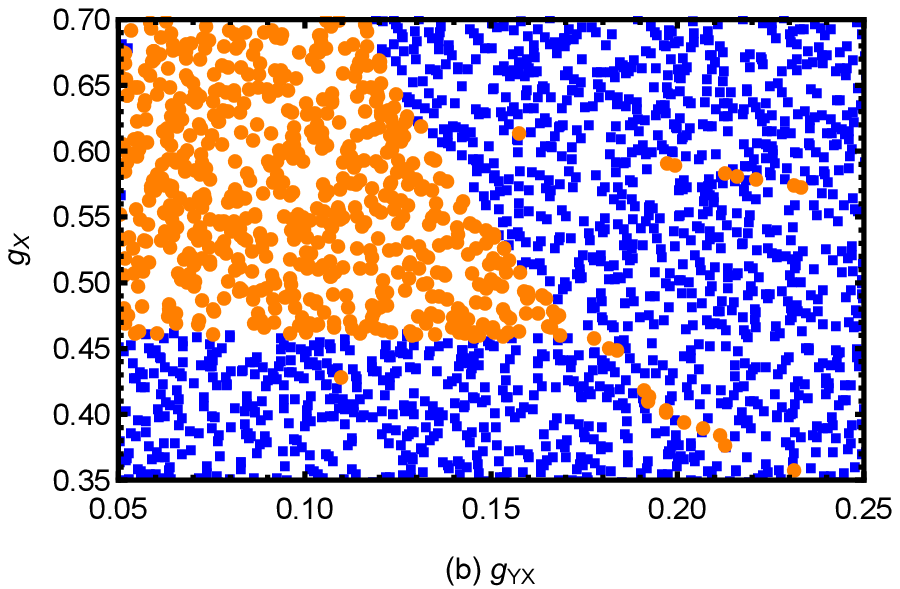}
\caption{${\Gamma}_{U(1)_X}(h\rightarrow \Upsilon Z )/{\Gamma}_{SM}(h\rightarrow \Upsilon Z)$ in $g_{X}-m^2_{\tilde{L}}$ plane (a) and $g_{YX}-g_{X}$ plane (b).}{\label {8}}
\end{figure}

\section{Conclusion}

The local gauge group of the $U(1)_X$SSM is $SU(3)_C\times SU(2)_L \times U(1)_Y\times U(1)_X$, and it is the expansion of MSSM. $U(1)_X$SSM has new superfields including right-handed neutrinos $\hat{v}_i$ and three Higgs superfields $\hat{\eta},~\hat{\bar{\eta}},~\hat{S}$. In the framework of $U(1)_X$SSM, we have performed a detailed analysis of the Higgs boson decays $h\rightarrow Z \gamma$ and $h\rightarrow m_V Z$, where $m_V$ denotes $\omega,\rho,\phi,J/{\psi}$ and $\Upsilon$. The numerical values of the processes $h\rightarrow \gamma \gamma$ and $h\rightarrow V V^*(V=Z,W)$ are discussed. $h ZZ$ coupling exists at the tree level in the SM, and the $h \gamma Z$ coupling is generated by loop diagrams. In models beyond SM, there can be CP-even coupling constant $C_{\gamma Z}$ and CP-odd coupling constant $\tilde{C}_{\gamma Z}$. The CP-even part is more important than the CP-odd part. There are direct and indirect contributions for the decay $h\rightarrow m_V Z$. The indirect contributions are more essential than the direct contributions. In general, the contributions of particles in the loops are inversely proportional to their mass square.
The heavier the particle mass, the smaller its contributions. Thus, sleptons give the most important contributions to the loop corrections and not other particles.

Using the effective Lagrangian method, we calculate the effective constants $C_{\gamma Z}$ and $\tilde{C}_{\gamma Z}$ for the vertex $h \gamma Z$. Furthermore, we obtain abundant numerical results by scanning parameter space. After analyzing and comparing these interesting graphics, we can conclude that large $\tan{\beta}_{\eta}$, small $m^2_{\tilde{L}}$ and large $g_{YX}$ have great impacts on the results. The insensitive parameters $\tan{\beta}$,~$T_e$ and $M_2$ mildly influence the numerical results. Comparing with the SM results, the new physics corrections to $h\rightarrow \gamma \gamma$ are around 1.1 and the new physics corrections to $h\rightarrow V V^*(V=Z,W)$ are in ($1.1-1.3$). The ratio ${\Gamma}_{U(1)_X}(h\rightarrow Z \gamma )/{\Gamma}_{SM}(h\rightarrow Z \gamma)$ is between $1.02-1.35$.
For the vector mesons $\omega,\rho,\phi$ and $J/{\psi}$, the ratios ${\Gamma}_{U(1)_X}(h\rightarrow m_V Z )/{\Gamma}_{SM}(h\rightarrow m_V Z)$ are mainly distributed in the range of ($ 1.01 - 1.45 $). When $m_V$  represents $\Upsilon$, the larger value of the ratio ${\Gamma}_{U(1)_X}(h\rightarrow \Upsilon Z )/{\Gamma}_{SM}(h\rightarrow \Upsilon Z)$ is mostly located in the range of ($ 1.002-1.30 $). We can find an interesting law: with the increase of the final state meson mass, new physics corrections of the $U(1)_X$SSM become small.

Through the above analysis, the decay $h\rightarrow m_V Z$ can be realized on HL-LHC and future high energy colliders. This work has certain reference value for detecting the decays $h\rightarrow Z \gamma$ and $h\rightarrow m_V Z$ and exploring new physics beyond SM. If the decay $h\rightarrow Z \gamma$  was measured and
its rate turned out to be larger than the SM prediction,
it will further confirm the interesting sensitivity of the rare Higgs boson decay in the $U(1)_X$SSM
and support the study of  $U(1)_X$SSM.

\begin{acknowledgments}

This work is supported by National Natural Science Foundation of China (NNSFC)
(No. 12075074), Natural Science Foundation of Hebei Province
(A2020201002, A202201022), Natural Science Foundation of Hebei Education Department (QN2022173), Post-graduate's Innovation Fund Project of Hebei University (HBU2022ss028), Post-graduate's Innovation Fund Project of Hebei (Hebei University) (Higgs boson decays $h\rightarrow Z \gamma$ and $h\rightarrow m_V Z$ in the $U(1)_X$SSM).
\end{acknowledgments}

\appendix
\section{Used coupling in $U(1)_X$SSM}

Here, we show some corresponding vertexes in this model. Their concrete forms are shown as
\begin{eqnarray}
&&\mathcal{L}_{Z H H}= \frac{1}{2}{H}^{+}_j
\Big(-g_1\cos\theta_W^\prime\sin\theta_W+g_2\cos\theta_W\cos\theta_W^\prime
+(g_{YX}+g_{X})\sin\theta_W^\prime\Big)\nonumber\\&&\hspace{1.4cm}
(\sum_{a=1}^3Z_{i,a}^{H,*}Z_{j,a}^H+\sum_{a=1}^3Z_{i,3+a}^{H,*}Z_{j,3+a}^H)
(p^\mu_{i}-p^\mu_j){H}^-_iZ_{\mu},
\end{eqnarray}
\begin{eqnarray}
&&\mathcal{L}_{H\tilde{L}\tilde{L}^{*}}=\frac{1}{4} \tilde{L}_i \Big\{ \sum_{a=1}^3Z_{j,a}^{E,*}Z_{i,a}^E \Big[(g^2_2-g_{YX} g_{X}-g^2_1-g^2_{YX}-4 Y^{2}_{e,a})v_d Z^H_{k1}\!+(\!-g^2_2\!+g_{YX} g_{X}\nonumber\\&&\hspace{1.2cm}+g^2_1+g^2_{YX})v_u Z^H_{k2}-2 g_{YX} g_{X}(v_{\eta} Z^H_{k3}-v_{\bar{\eta}} Z^H_{k4}) \Big] +\sum_{a=1}^3Z_{j,3+a}^{E,*}Z_{i,3+a}^E \Big[(2 g^2_1+2 g^2_{YX}\nonumber\\&&\hspace{1.2cm} +3g_{YX} g_{X}+g^2_{X}-4 Y^{2}_{e,a})v_d Z^H_{k1}-(2 g^2_1+2g^2_{YX}+3g_{YX} g_{X}+g^2_{X})v_u Z^H_{k2}\nonumber\\&&\hspace{1.2cm}+(4 g_{YX} g_{X}\!+\!2 g^2_{X})(v_{\eta} Z^H_{k3}\!-\!v_{\bar{\eta}} Z^H_{k4})\Big]
\!+\!\sum_{a=1}^3Z_{j,a}^{E,*}Z_{i,3+a}^E \Big[\!-\!2\sqrt{2} T_{e,a} Z^H_{k1}\!+\!2(v_S {\lambda}^{*}_{H} Y_{e,a}\nonumber\\&&\hspace{1.2cm}+\sqrt{2}{\mu}^{*}Y_{e,a})Z^H_{k2}+2 v_u {\lambda}^{*}_{H}Y_{e,a}Z^H_{k5}\Big]
\!+\!\sum_{a=1}^3Z_{j,3+a}^{E,*}Z_{i,a}^E\Big[-2\sqrt{2} T^{*}_{e,a} Z^H_{k1}+2(v_S {\lambda}_{H} Y^{*}_{e,a}\nonumber\\&&\hspace{1.2cm}+\sqrt{2}{\mu}Y^{*}_{e,a})Z^H_{k2}+2 v_u {\lambda}_{H}Y^{*}_{e,a} Z^H_{k5}\Big] \Big\}\tilde{L}_j^{*} H_k,
\end{eqnarray}
\begin{eqnarray}
&&\mathcal{L}_{H \tilde{D} \tilde{D}}=\frac{1}{12}\tilde{D}_i \Big\{ \sum_{a=1}^3Z_{j,a}^{D,*}Z_{i,a}^D \Big[(3 g^2_2+g_{YX} g_{X}+g^2_1+g^2_{YX}\!-\!12 Y^{2}_{d,a})v_d Z^H_{k1}-(3 g^2_2+g_{YX} g_{X}
\nonumber\\&&\hspace{1.2cm}+g^2_1+g^2_{YX})v_u Z^H_{k2}
+2 g_{YX} g_{X}(v_{\eta} Z^H_{k3}-v_{\bar{\eta}} Z^H_{k4}) \Big]
+\sum_{a=1}^3Z_{j,3+a}^{D,*}Z_{i,3+a}^D \Big[(2 g^2_1+2 g^2_{YX}
\nonumber\\&&\hspace{1.2cm} +3 g^2_{X}+5g_{YX} g_{X}-12 Y^{2}_{d,a})v_d Z^H_{k1}-(2 g^2_1+2g^2_{YX}+3 g^2_{X}+5g_{YX} g_{X})v_u Z^H_{k2}
\nonumber\\&&\hspace{1.2cm}+(4 g_{YX} g_{X}+6 g^2_{X})(v_{\eta} Z^H_{k3}-v_{\bar{\eta}} Z^H_{k4})\Big]+ \sum_{a=1}^3Z_{j,a}^{D,*}Z_{i,3+a}^D \Big[(-6\sqrt{2} T_{d,a} Z^H_{k1})+6(v_S {\lambda}^{*}_{H} Y_{d,a}
\nonumber\\&&\hspace{1.2cm}+\sqrt{2}{\mu}^{*}Y_{d,a})Z^H_{k2}+6 v_u {\lambda}^{*}_{H}Y_{d,a}Z^H_{k5}\Big]+\sum_{a=1}^3Z_{j,3+a}^{D,*}Z_{i,a}^D\Big[(-6\sqrt{2} T^{*}_{d,a} Z^H_{k1})+6(v_S {\lambda}_{H} Y^{*}_{d,a}
\nonumber\\&&\hspace{1.2cm}+\sqrt{2}{\mu}Y^{*}_{d,a})Z^H_{k2}+6 v_u {\lambda}_{H}Y^{*}_{d,a} Z^H_{k5}\Big] \Big\}\tilde{D}_j^{*} H_k,
\end{eqnarray}
\begin{eqnarray}
&&\mathcal{L}_{H \tilde{U} \tilde{U}}=\frac{1}{12}\tilde{U}_i\Big\{ \sum_{a=1}^3Z_{j,a}^{U,*}Z_{i,a}^U \Big[ (-3 g^2_2+g_{YX} g_{X}+g^2_1+g^2_{YX})v_d Z^H_{k1}-(-3 g^2_2+g_{YX} g_{X}+g^2_1
\nonumber\\&&\hspace{1.1cm} +g^2_{YX}-12 Y^{2}_{u,a})v_u Z^H_{k2}
+2 g_{YX} g_{X}(v_{\eta} Z^H_{k3}-v_{\bar{\eta}} Z^H_{k4}) \Big]+\sum_{a=1}^3Z_{j,3+a}^{U,*}Z_{i,3+a}^U  \Big[-(3 g^2_{X}
\nonumber\\&&\hspace{1.1cm}+4 g^2_1+4 g^2_{YX}+7 g_{YX} g_{X})v_d Z^H_{k1}+(3 g^2_{X}+4 g^2_1+4 g^2_{YX}+7 g_{YX} g_{X}-12 Y^{2}_{u,a})v_u Z^H_{k2}
\nonumber\\&&\hspace{1.1cm}-(8 g_{YX} g_{X}+6 g^2_{X})
(v_{\eta} Z^H_{k3}-v_{\bar{\eta}} Z^H_{k4}) \Big]
+ \sum_{a=1}^3Z_{j,a}^{U,*}Z_{i,3+a}^U \Big[6(v_S {\lambda}^{*}_{H} Y_{u,a}+\sqrt{2}{\mu}^{*}Y_{u,a})Z^H_{k1})
\nonumber\\&&\hspace{1.1cm} -6\sqrt{2} T_{u,a} Z^H_{k2}+6 v_d {\lambda}^{*}_{H}Y_{u,a}Z^H_{k5}\Big]+ \sum_{a=1}^3Z_{j,3+a}^{U,*}Z_{i,a}^U \Big[6(v_S {\lambda}_{H} Y^{*}_{u,a}+\sqrt{2}{\mu}Y^{*}_{u,a})Z^H_{k1})
\nonumber\\&&\hspace{1.1cm} -6\sqrt{2} T^{*}_{u,a} Z^H_{k2}+6 v_d {\lambda}_{H}Y^{*}_{u,a}Z^H_{k5}\Big] \Big\}\tilde{U}_j^{*} H_k.
\end{eqnarray}

\end{document}